\documentclass[aps,prl,letterpaper,twocolumn,preprintnumbers,floatfix,superscriptaddress]{revtex4}

\usepackage{mathrsfs}
\DeclareMathAlphabet{\mathpzc}{OT1}{pzc}{m}{it}

\usepackage[english]{babel}

\usepackage{amsmath}
\usepackage{amssymb}

\usepackage{mathrsfs}
\DeclareMathAlphabet{\mathpzc}{OT1}{pzc}{m}{it}

\usepackage{graphicx}

\usepackage{cancel} 
\usepackage{paralist}
\usepackage{booktabs}
\usepackage{hyperref}
\usepackage{xspace}

\usepackage{graphicx,color,verbatim,acronym,epsfig,subfigure}
\usepackage{slashed,skak}
\usepackage{latexsym,amsfonts,amssymb,amsmath,makeidx,mathrsfs,multirow,lineno,bm,bbm}

\begin{document}
\title{On the gravitational wave  sources from the NANOGrav 12.5-yr data}

\author{Ligong Bian }\email{lgbycl@cqu.edu.cn}
\affiliation{Department of Physics, Chongqing University, Chongqing 401331, China}	

\author{Rong-Gen Cai}\email{cairg@itp.ac.cn}
\affiliation{CAS Key Laboratory of Theoretical Physics, Institute of Theoretical Physics, Chinese Academy of Sciences, P.O. Box 2735, Beijing 100190, China}
\affiliation{School of Physical Sciences, University of Chinese Academy of Sciences, No.19A Yuquan Road, Beijing 100049, China}
\affiliation{School of Fundamental Physics and Mathematical Sciences, Hangzhou Institute for Advanced Study, University of Chinese Academy of Sciences, Hangzhou 310024, China}

\author{Jing Liu}\email{liujing@ucas.ac.cn}
\affiliation{School of Fundamental Physics and Mathematical Sciences, Hangzhou Institute for Advanced Study, University of Chinese Academy of Sciences, Hangzhou 310024, China}
\affiliation{School of Physical Sciences, University of Chinese Academy of Sciences, No.19A Yuquan Road, Beijing 100049, China}

\author{Xing-Yu Yang}\email{yangxingyu@itp.ac.cn}
\affiliation{CAS Key Laboratory of Theoretical Physics, Institute of Theoretical Physics, Chinese Academy of Sciences, P.O. Box 2735, Beijing 100190, China}
\affiliation{School of Physical Sciences, University of Chinese Academy of Sciences, No.19A Yuquan Road, Beijing 100049, China}

\author{Ruiyu Zhou }\email{zhoury@cqu.edu.cn}
\affiliation{Department of Physics, Chongqing University, Chongqing 401331, China}

\date{\today}

\begin{abstract}

The NANOGrav Collaboration recently reported a strong evidence for a stochastic common-spectrum process in the
pulsar-timing data. We evaluate the evidence of interpreting this process as mergers of super massive black hole binaries and/or various stochastic gravitational wave background sources in the early Universe, including first-order phase transitions, cosmic strings, domain walls, and large amplitude curvature perturbations. We discuss the implications of the constraints on these possible sources. It is found that the cosmic string is the most favored source against other gravitational wave sources based on the Bayes factor analysis.

\end{abstract}

\graphicspath{{figure/}}

\maketitle
\baselineskip=16pt

\pagenumbering{arabic}

%

\noindent{\it \bfseries  Introduction:}
Recently, with the 12.5-yr data set, the NANOGrav Collaboration reports a strong evidence of a stochastic common-spectrum process~\cite{Arzoumanian:2020vkk}. The process may be interpreted as gravitational waves (GWs) from mergers of super massive black hole binaries (SMBHBs)~\cite{Rajagopal:1994zj,Phinney:2001di,Jaffe:2002rt,Wyithe:2002ep} with 
 spectral index of $13/3$~\cite{Arzoumanian:2020vkk}, or other speculative stochastic gravitational wave backgrounds (SGWBs) in the nanohertz frequency region, such as, cosmic strings~\cite{Siemens:2006yp}, first-order phase transition(FOPT)~\cite{Kosowsky:1992rz,Caprini:2010xv}, a primordial
GW background produced by quantum fluctuations of the gravitational field during inflation~\cite{Grishchuk:1974ny}, and domain walls~\cite{Hiramatsu:2013qaa}.

Motivated by the observation, there are several attempts to interpret the result as GWs from
super massive black holes~\cite{Vaskonen:2020lbd,DeLuca:2020agl,Kohri:2020qqd}, cosmic strings~\cite{Ellis:2020ena,Blasi:2020mfx,Samanta:2020cdk}, dark phase transition~\cite{Nakai:2020oit,Addazi:2020zcj,Ratzinger:2020koh}, and large scalar fluctuations associated with primordial black hole (PBH) formation~\cite{DeLuca:2020agl,Vaskonen:2020lbd}. In this letter, based on the assumption that the stochastic common-spectrum process is attributed to GWs, we firstly apply a Bayesian analysis to evaluate the strength of evidence for explaining the stochastic common-spectrum process with different models, including: SMBHBs, cosmic strings, scalar perturbations, FOPT, and domain walls, and/or a superposition of some of these GW sources above. We find out the constraints on parameter spaces of these GWs models. The constraints on the gravitational wave parameters from all these sources can be used to limit:  1) the phase transition parameters of dark sector;  2) the symmetry breaking scale of new physics beyond the Standard Model of particle physics; and 3) the PBH abundance.  
Model comparison of Bayesian fit are performed to judge GW model interpretations with the current NANOGrav 12.5-yr data.

\noindent{\it \bfseries  NANOGrav 12.5-yr Results versus GW sources:}
We perform a Bayesian inference over the first five-frequency bins of the NANOGrav 12.5-yr data set~\cite{Arzoumanian:2020vkk}, roughly $f \in (2.5\times 10^{-9},1.2\times 10^{-8})$~Hz. 
We apply full Bayesian model fitting and model comparison procedure to these explanations.
The free parameters of each model are set with uniform priors. 

 {\it I)  GWs from FOPTs:}
Two crucial parameters for the calculation of GWs from FOPT are: 1) the latent heat~\cite{Caprini:2015zlo}, $\alpha_{PT}$; 2) the inverse time duration of the phase transition $
\frac{\beta}{H_n}$, where $T_{n}$ is the PT temperature and $H_{n}$ is the Hubble parameter.
We consider 
the nonrun-away bubbles where the GW sources from FOPT are dominated by sound waves, and the GW energy spectrum reads~\cite{Caprini:2015zlo}
\begin{eqnarray}
\Omega_{\mathrm{GW}}^{\rm sw} h^2(f)&=&2.65 \times 10^{-6}(H_*\tau_{sw})\left(\frac{\beta}{H_*}\right)^{-1} v_b
\left(\frac{\kappa_\nu \alpha_{PT} }{1+\alpha_{PT} }\right)^2\nonumber
\\
&\times&\left(\frac{g_*}{100}\right)^{-\frac{1}{3}}
\left(\frac{f}{f_{\rm sw}}\right)^3 \left(\frac{7}{4+3 \left(f/f_{\rm sw}\right)^2}\right)^{7/2}\,.
\end{eqnarray}
Here, the factor $\tau_{sw}={\rm min} \left[\frac{1}{H_*},\frac{R_*}{\bar{U}_f}\right]$ is adopted to account for the duration of the phase transition~\cite{Ellis:2019oqb,Ellis:2020awk,Caprini:2019egz,Ellis:2018mja,Guo:2020grp}, where $H_*R_*=v_b(8\pi)^{1/3}(\beta/H_*)^{-1}$~\cite{Ellis:2020awk}, and the root-mean-square (RMS) fluid velocity is
$
\bar{U}_f^2\approx\frac{3}{4}\frac{\kappa_\nu\alpha}{1+\alpha}
$~\cite{Hindmarsh:2017gnf, Caprini:2019egz, Ellis:2019oqb}.
The factor $\kappa_\nu$ is the fraction of the latent heat transferred into the kinetic energy of plasma, which is obtained by the hydrodynamic analysis~\cite{Espinosa:2010hh}.
The peak frequency of sound waves locates at
$
f^{\rm sw}=1.9 \times 10^{-5} \frac{\beta}{H} \frac{1}{v_b} \frac{T_*}{100}\left({\frac{g_*}{100}}\right)^{\frac{1}{6}} {\rm Hz }
$~\cite{Hindmarsh:2013xza,Hindmarsh:2015qta,Hindmarsh:2017gnf}.
For this study, we consider the plasma temperature being $T_\star\approx T_n$. 

\begin{figure}[!htp]
\begin{center}
\includegraphics[width=0.15\textwidth]{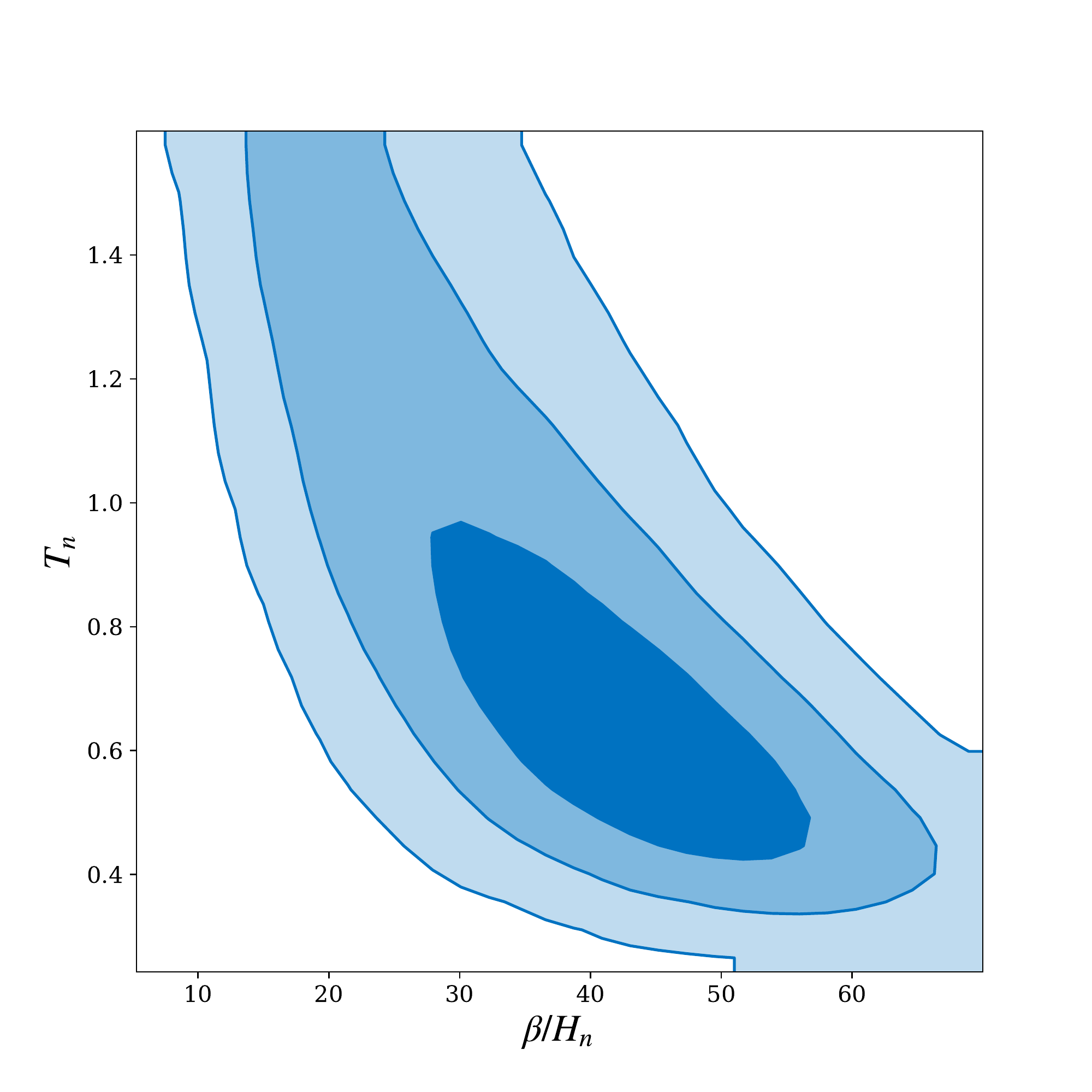}
\includegraphics[width=0.15\textwidth]{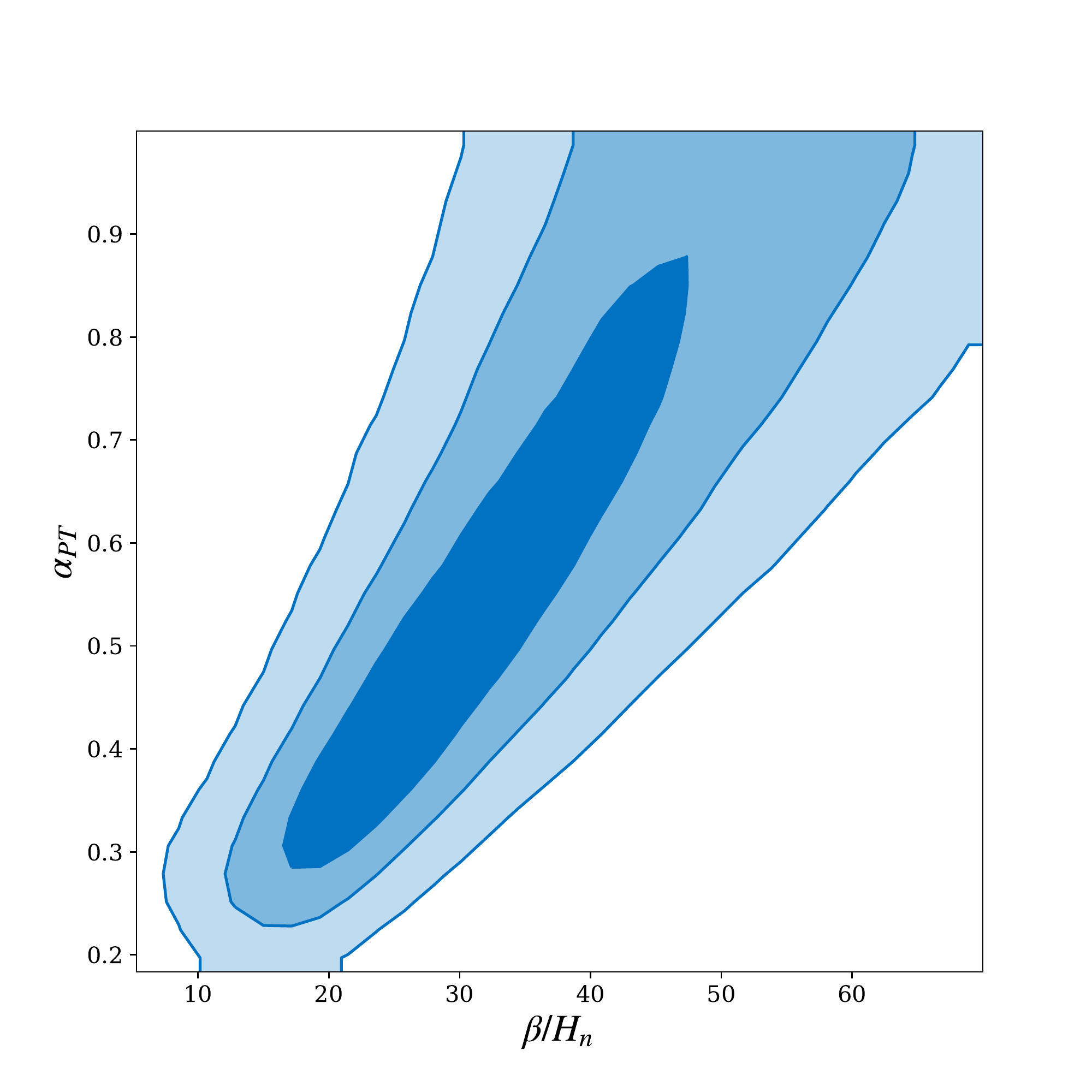}
\includegraphics[width=0.15\textwidth]{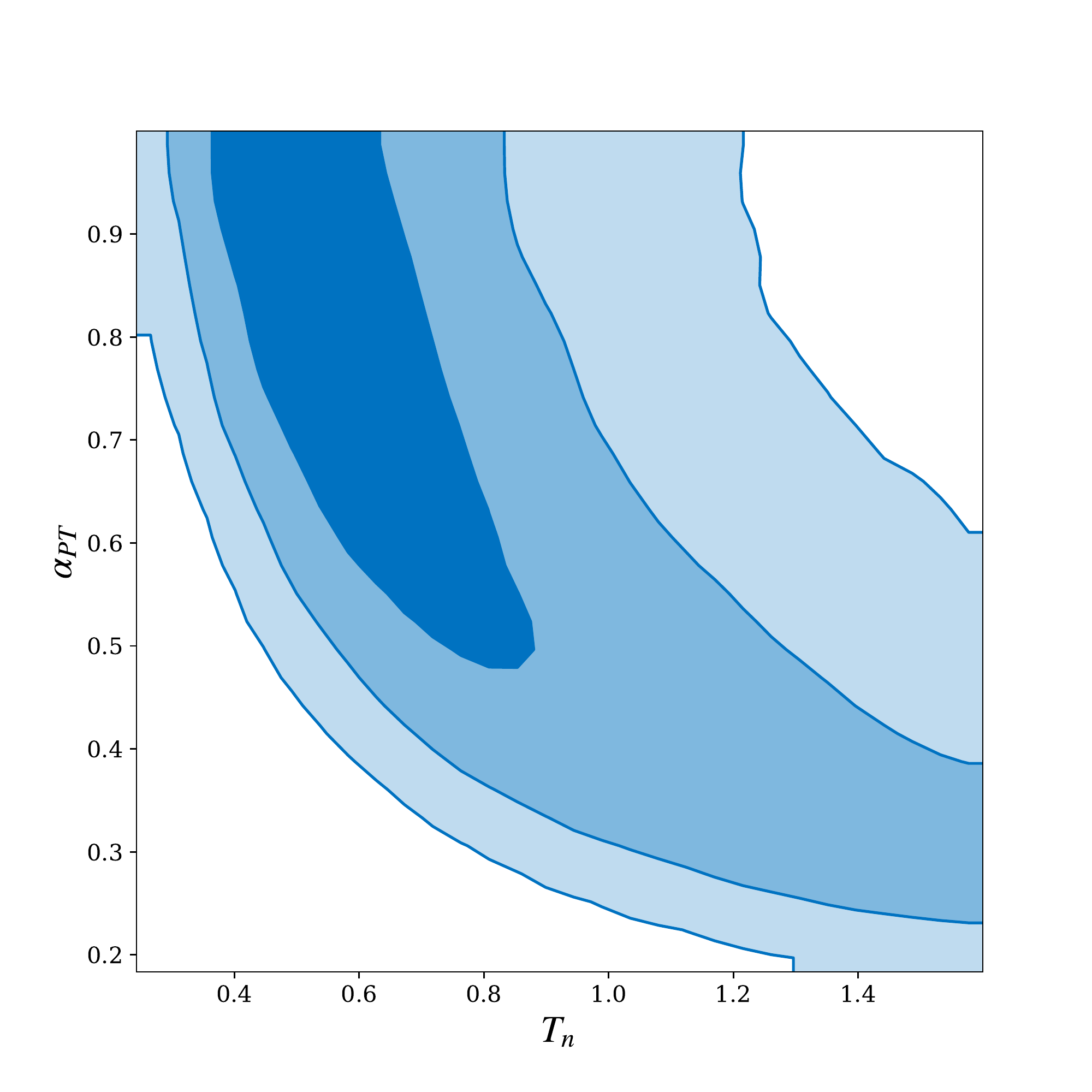}
\includegraphics[width=0.15\textwidth]{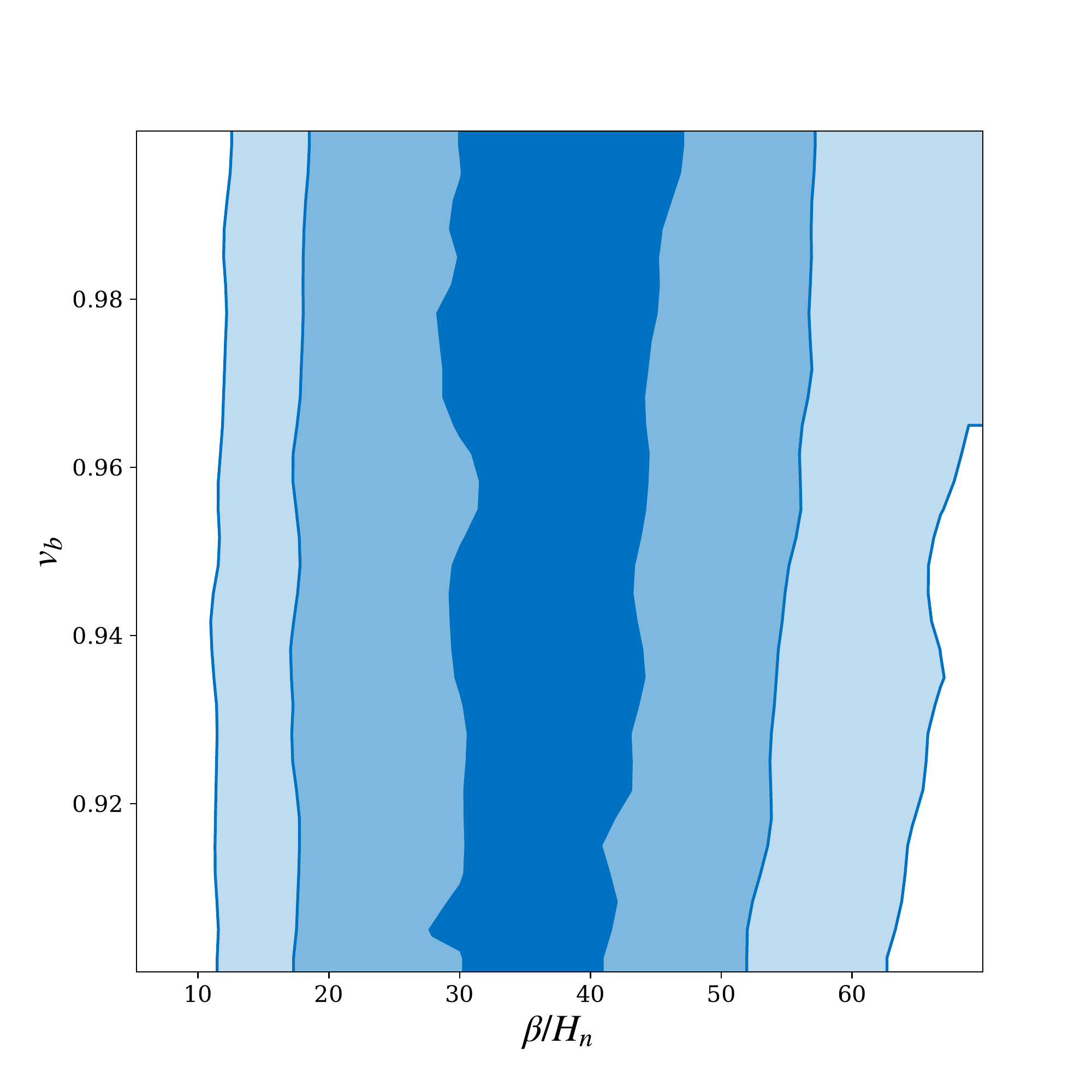}
\includegraphics[width=0.15\textwidth]{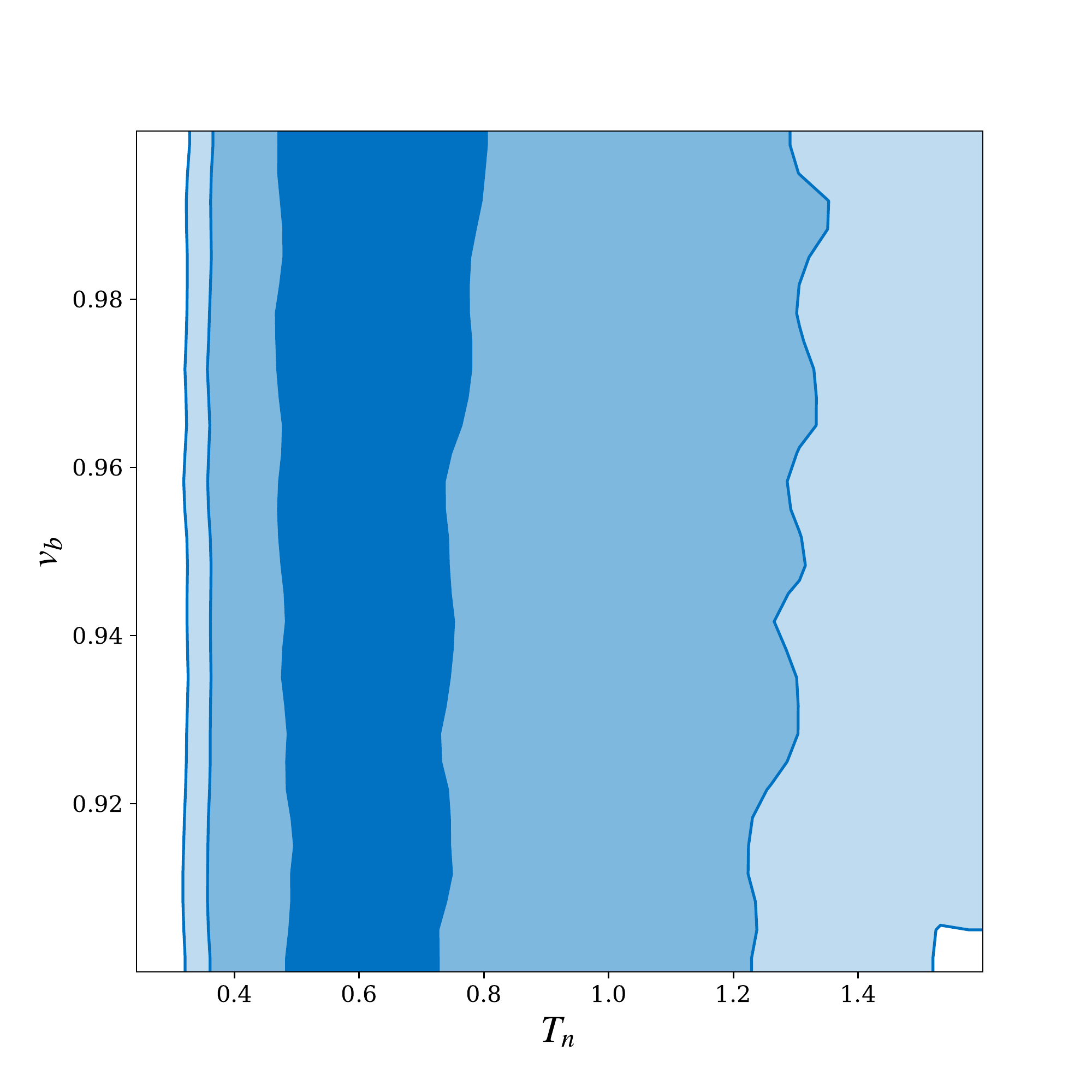}
\includegraphics[width=0.15\textwidth]{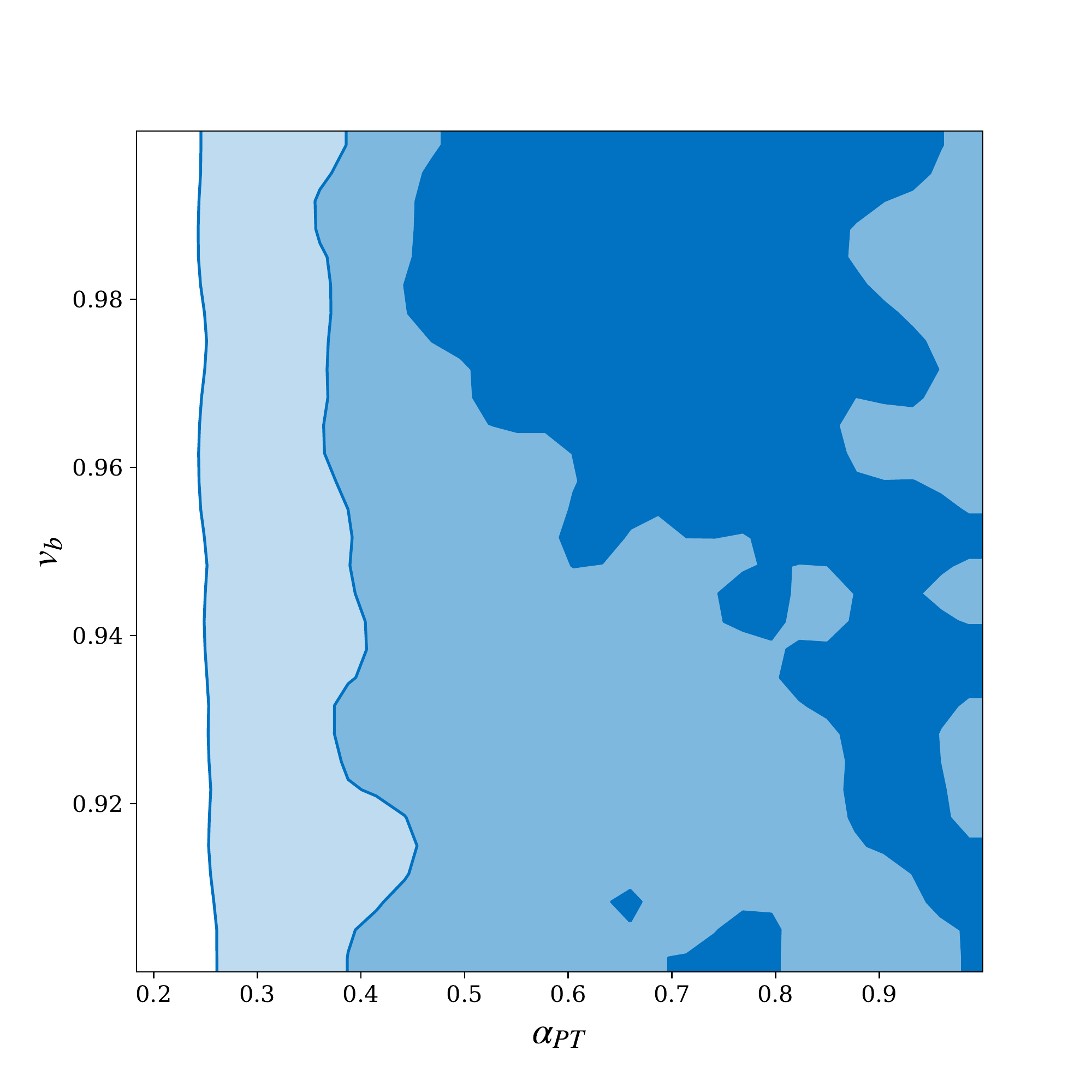}
\caption{The constraints on parameters of FOPT from Bayesain model fitting.
 The contour levels in the plots correspond to 1-, 2- and 3-$\sigma$ level in 2D distributions. }\label{PTGW}
\end{center}
\end{figure}

In Fig.~\ref{PTGW}, we perform a whole parameter space scan, the data constraints at 1 $\sigma$ favors: a supersonic velocity $v_b\sim[0.91,0.99]$, a moderate latent heat $\alpha_{PT}\sim [0.31,0.96]$, with a duration $\beta/H_{n}\sim [18.17,58.27]$ at phase transition temperature $T_n~\sim [0.42,1.43]$ MeV for FOPT. We note that the BBN bounds, $T_\star
\geq 1$ MeV~\cite{Breitbach:2018ddu}. This result may invalidate a lot parameter spaces of the dark phase transition explanation of the NANOGrav observation~\cite{Nakai:2020oit,Addazi:2020zcj,Ratzinger:2020koh} at this level. 

 {\it II) GWs from cosmic strings:}
The vastly adopted Nambu-Goto cosmic strings are characterized solely by the dimensionless parameter $G\mu$,  where $G$ is Newton's constant and $\mu$ is the string tension which is a function of the breaking scale of the $U(1)$ symmetry. We consider  GWs emitted by cosmic string network is dominated by cusps here, with~\cite{Cui:2018rwi}
$
\Omega_{\rm GW}^{cs}(f) =\sum_k \Omega_{\rm GW}^{(k)}(f)\; $,
and for each $k$-mode~\footnote{In Ref.~\cite{Gouttenoire:2019rtn,Gouttenoire:2019kij}, one can find the case with thermal frictions and particle productions.} 
\begin{eqnarray}\label{eq:GWdensity2}
\Omega_{\rm GW}^{(k)}(f) =&&
\frac{1}{\rho_c}
\frac{2k}{f}
\frac{\mathcal{F}_{\alpha}\,\Gamma^{(k)}G\mu^2}
{\alpha_{CS}\left( \alpha_{CS}+\Gamma G\mu\right)}
\int_{t_F}^{t_0}\!d\tilde{t}\;
\frac{C_{eff}(t_i^{(k)})}{t_i^{(k)\,4}}\nonumber\\
&&\times\bigg[\frac{a(\tilde{t})}{a(t_0)}\bigg]^5\bigg[\frac{a(t^{(k)}_i)}{a(\tilde{t})}\bigg]^3\Theta(t_i^{(k)} - t_F)~.~~~~
\end{eqnarray}
Here $\rho_c = 3H_0^2/8\pi G$ is the critical density, and
we take $\mathcal{F}_{\alpha}=0.1$ to characterize the fraction of the energy released by long strings. The loop production efficiency is adopted as $C_{eff}=5.4(0.39)$ in the radiation (matter) dominated universe~\cite{Gouttenoire:2019kij}.
The gravitational loop-emission efficiency is $\Gamma\approx50$~\cite{Blanco-Pillado:2017oxo}. The fourier modes of cusps~\cite{Olmez:2010bi}
are~\cite{Blanco-Pillado:2013qja,Blanco-Pillado:2017oxo}:
$
\Gamma^{(k)} = \frac{\Gamma k^{-\frac{4}{3}}}{\sum_{m=1}^{\infty} m^{-\frac{4}{3}} } \;,
$
here $\sum_{m=1}^{\infty} m^{-\frac{4}{3}} \simeq 3.60$ and
$\sum_k \Gamma^{(k)}=\Gamma$.
The formation time of loops of the $k$-mode casts the form of
\begin{equation}\label{eq:ti}
t_i^{(k)}(\tilde{t},f) = \frac{1}{\alpha_{CS}+\Gamma G\mu}\left[
\frac{2 k}{f}\frac{a(\tilde{t})}{a(t_0)} + \Gamma G\mu\;\tilde{t}\;
\right]\,,
\end{equation}
where $\tilde{t}$ is the GW emission time. The cosmic string network reaches an attractor scaling solution after the formation time, $t_F$. When the small-scale structure of loops is dominated by cusps, the high mode relates to the low mode as:
$\Omega_{\rm GW}^{(k)}(f)
= \frac{\Gamma^{(k)}}{\Gamma^{(1)}}\,\Omega_{\rm GW}^{(1)}(f/k)
=k^{-4/3}\,\Omega_{\rm GW}^{(1)}(f/k)$\;.

\begin{figure}[!htp]
\begin{center}
\includegraphics[width=0.4\textwidth]{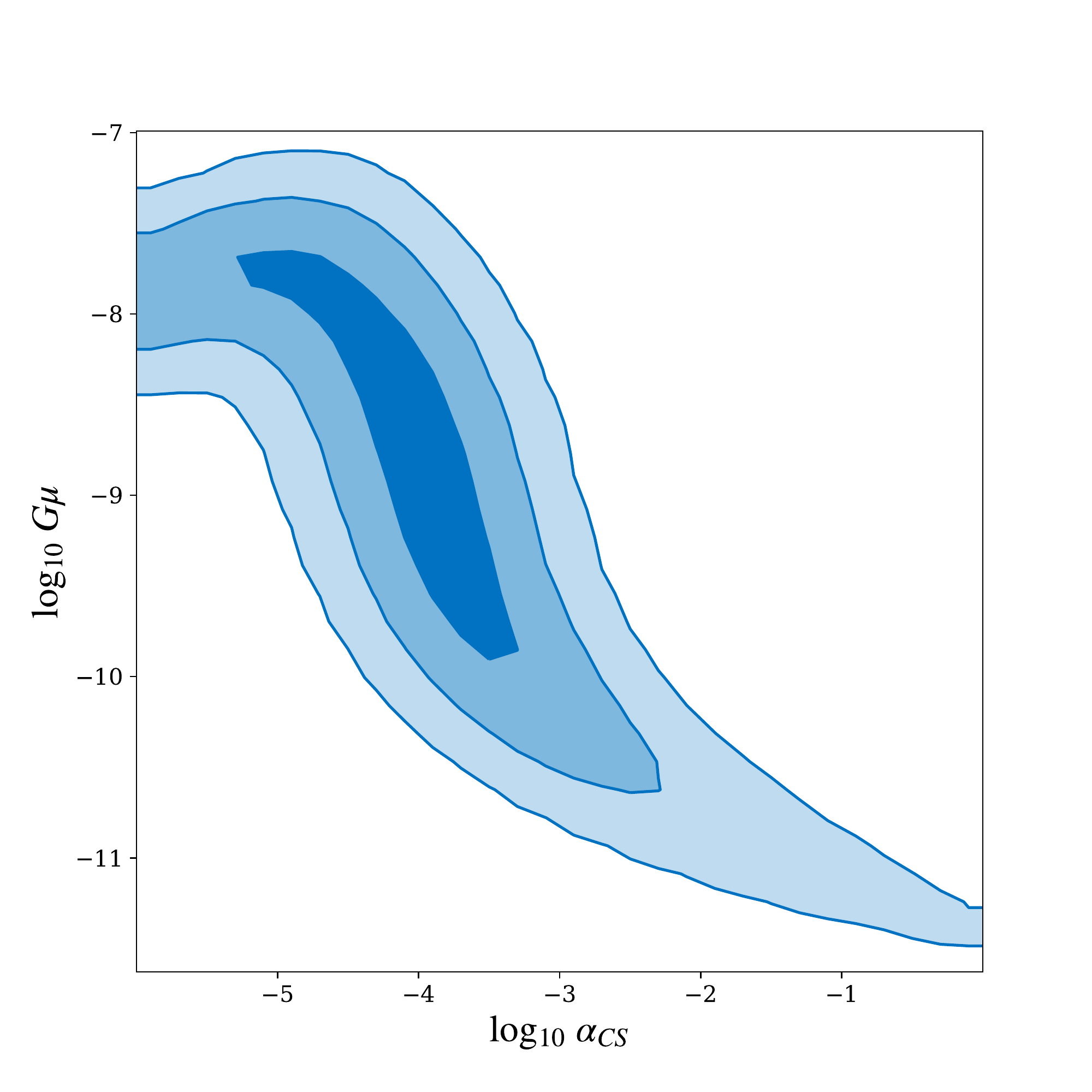}
\caption{ The constraints on parameters of cosmic strings from the  Bayesian model fitting.}
\label{CSGW}
\end{center}
\end{figure}

In Fig.~\ref{CSGW} we show the results of Bayesian model fitting for the cosmic string network case. The constraints yields $\log_{10} G\mu\sim[-10.44, -7.64]$ at 1 $\sigma$, which suggests the U(1) symmetry breaking scale of the new physics connecting with grant unification is around $\eta~\sim \mathcal{O}(10^{14-15})$ GeV) for local strings (where $\mu\approx 2 \pi \eta^2$ for the winding number $n=1$).  
 And $\log_{10} \alpha_{CS}\sim [-3.91,-2.63]$
is allowed by data at 1 $\sigma$,  which is slightly smaller than the $\alpha_{CS}=0.1$ suggested by the simulations~\cite{Blanco-Pillado:2013qja,Blanco-Pillado:2017oxo}.

{\it III) GWs from the domain wall decay:} The peak amplitude and the frequency of GWs are determined by the surface energy density $\sigma$ (which is a function of the discrete symmetry scale) and the 
bias term $\Delta V$  (which explicitly break the discrete symmetry and is bound bellow since domain walls should decay before domination). 
We use the slope of spectrum $\Omega_{\mathrm{GW}}^{dw}h^2 \propto f^3$ when $f \textless f_{\rm peak}$, and $\Omega^{dw}_{\mathrm{GW}}h^2 \propto f^{-1}$ when $f \geqslant f_{\rm peak}$~as suggested by the estimation of Ref.~\cite{Hiramatsu:2013qaa}. 
The peak amplitude of the GW is~\cite{Hiramatsu:2013qaa,Kadota:2015dza,Zhou:2020ojf}:
\begin{eqnarray}
\Omega^{\mathrm{peak}}_{\mathrm{GW}} h^{2} &\simeq& 5.20 \times 10^{-20} \times \tilde{\epsilon}_{\mathrm{gw}} \mathcal{A}^{4}\left(\frac{10.75}{g_{*}}\right)^{1 / 3}\left(\frac{\sigma}{1 \mathrm{TeV}^{3}}\right)^{4}\nonumber\\
&&
\times\left(\frac{1 \mathrm{MeV}^{4}}{\Delta V}\right)^{2}\,,\label{eq:gwdw}
\end{eqnarray}
 with the peak frequency~\cite{Hiramatsu:2013qaa}:
$f^{dw}\left(t_{0}\right)_{\mathrm{peak}}\simeq 3.99 \times 10^{-9} \mathrm{Hz} \mathcal{A}^{-1 / 2}\left(\frac{1 \mathrm{TeV}^{3}}{\sigma}\right)^{1 / 2}\left(\frac{\Delta V}{1 \mathrm{MeV}^{4}}\right)^{1 / 2}\;$.
The area parameter $\mathcal{A}=1.2$~\cite{Kadota:2015dza}, the efficiency parameter $\tilde{\epsilon}_{\mathrm{gw}}=0.7$~\cite{Hiramatsu:2013qaa}, and the effective relativistic degree of freedom at the domain wall decay time $g_{*}= 10.75$ ~\cite{Kadota:2015dza}.

\begin{figure}[!htp]
\begin{center}
\includegraphics[width=0.4\textwidth]{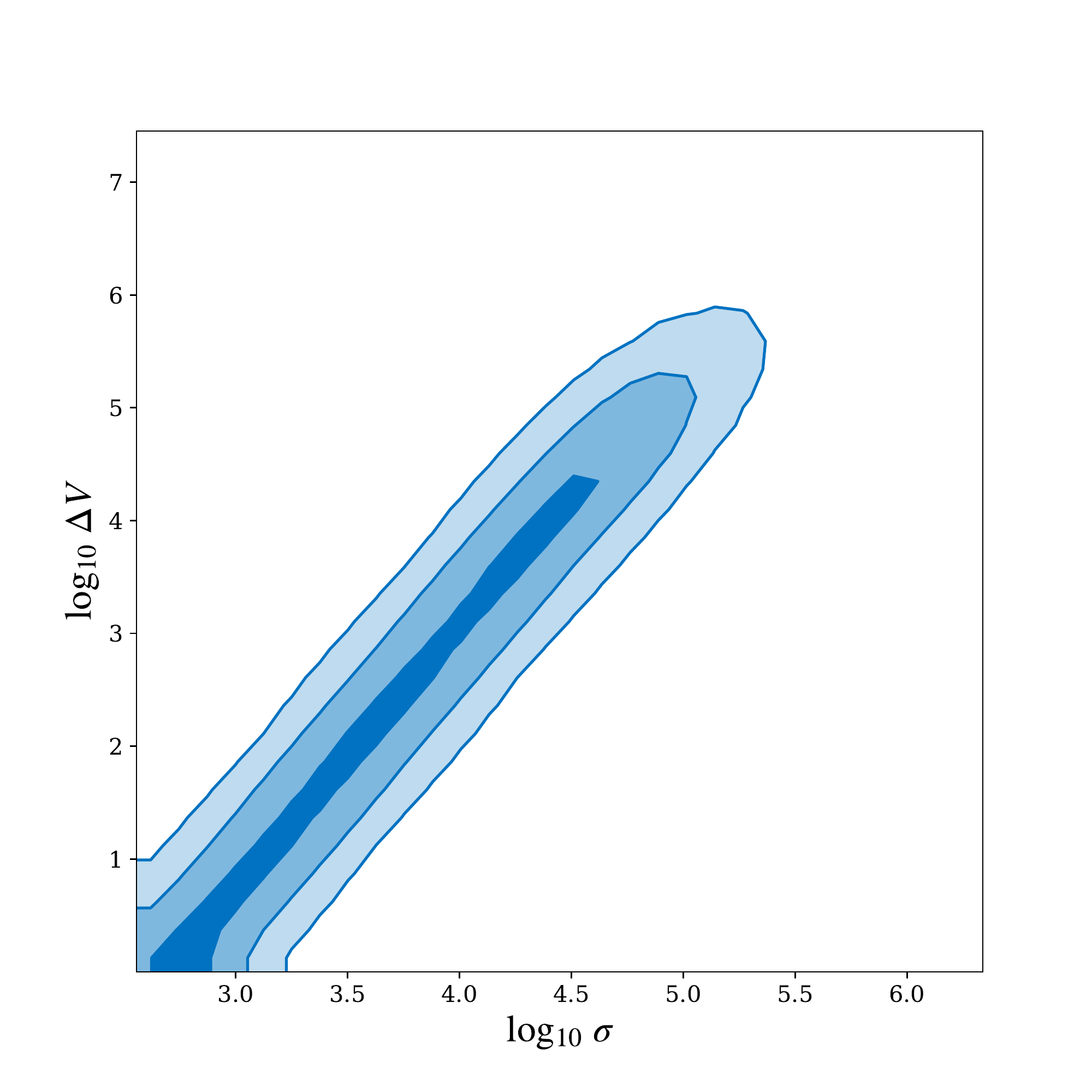}
\caption{The constraints on parameters of domain wall from Bayesain model fitting.  }
\label{DWGW}
\end{center}
\end{figure}

A higher magnitude of GWs is obtained with a large surface energy density. The Bayesian fit at 1 $\sigma$ sets the bound on the bias term and the surface energy density as: $\log_{10} (\sigma/\mathrm{TeV}^{3})\sim[2.79, 4.83]$,
        $\log_{10} (\Delta V/\mathrm{MeV}^{4})\sim[0.26,4.89]$. 
Utilizing $\sigma\sim 2\sqrt{2\lambda} \eta^3/3$ (here $\lambda$ and $\eta$ are the interaction coupling and the breaking scale for the $Z_2$ discrete symmetry), one can estimate the breaking scale of the discrete symmetry being $\eta \lesssim 10^{2}$ TeV for $\lambda\sim  \mathcal{O}(10^{-2})$.

{\it IV) Scalar-induced GWs:} 
 We use the method  in Ref.~\cite{Kohri:2018awv} to calculate the energy spectrum of the SGWB  resulting from the large scalar fluctuations during inflation, and study the constraints from the common-spectrum process detected by NANOGrav.
The GW production process happens almost around the horizon reentry of the corresponding modes, after that $\Omega_{\mathrm{GW}}$ soon reaches a constant. Assuming a power-law form of  scalar fluctuations $P_{\mathcal{R}}(k)=P_{\mathcal{R}0}k^{m}$ around $f=1\mathrm{yr}^{-1}$,
one can simply obtain $\Omega_{\mathrm{GW}}(t_0,f=1\mathrm{yr}^{-1})\propto P_{\mathcal{R}0}^{2}$ and $\Omega_{\mathrm{GW}}(k)\propto k^{2m}$ from the quadratic $P_{\mathcal{R}}$-dependence of $\Omega_{\mathrm{GW}}(t_{0},k)$.

\begin{figure}[!htp]
	\begin{center}
		\includegraphics[width=0.4\textwidth]{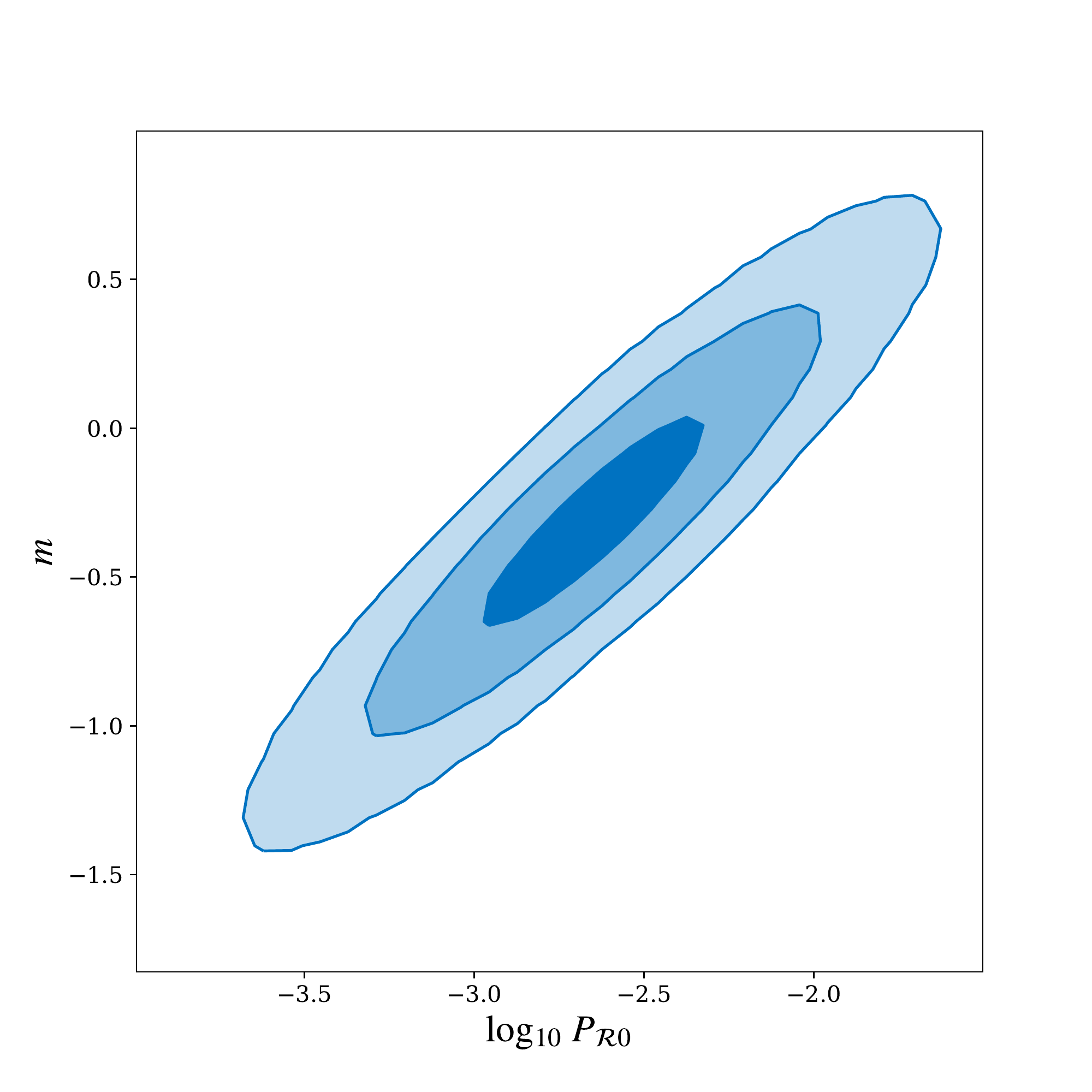}
		\caption{The constraints on parameters of power spectrum of curvature perturbations from the Bayesian model fitting.  }
        \label{IGW}
	\end{center}
\end{figure}

In the case of scalar-induced GWs, we find $\log_{10}P_{\mathcal{R} 0} \sim\left[-3.20, -2.10\right]$ and $m \sim[-0.89,0.27]$ are allowed
by the NANOGrav data at 1 $\sigma$, as
shown in Fig.~\ref{IGW}. This shows that the universal behavior in the  low-frequency region with $k^{3}$ slope \cite{Cai:2019cdl} is highly disfavored by the NANOGrav data. It is well-known that large amplitude
scalar perturbations are also responsible for the production of PBHs, which may constitute dark matter and provide merger events of black hole binaries ~\cite{Sasaki:2016jop,Sasaki:2018dmp,Carr:2020gox}. The
best-fit $P_{\mathcal{R}}$ of Gaussian curvature perturbations hints that the abundance of such PBHs is less than $10^{-12}$ for the PBH mass about $0.1 M_{\odot}$. The constraint on PBH mass function from NANOGrav is much more strict than the others in the same mass range~(less than $0.01$) such as microlensing results from EROS/MACHO~\cite{Tisserand:2006zx} and OLGE~\cite{Niikura:2019kqi}. However, if we consider non-Gaussian scalar perturbations
reported in Refs.~\cite{Cai:2018dig,Bartolo:2018rku,Cai:2019elf}, the $\mathrm{PBH}$ abundance can be
$10^{-3}$ to explain the merger rate observed by LIGO. The amplitude of scalar induced GWs is also in consistent with the assumption that PBH could seed the super massive black holes~\cite{Vaskonen:2020lbd}.  With the amplitude of scalar perturbations extended into smaller scales, the predicted PBH mass decreases and the PBH abundance might increase enormously. PBHs could constitute all dark matter when assuming a scale-invariant power spectrum for scalar fluctuations~\cite{Domenech:2020ers}, or explain the microlensing events observed by OGLE when taking into account the early dust-like stage~\cite{Domenech:2020ers}.

\begin{figure}[!htp]
    \centering
    \includegraphics[width=0.2\textwidth]{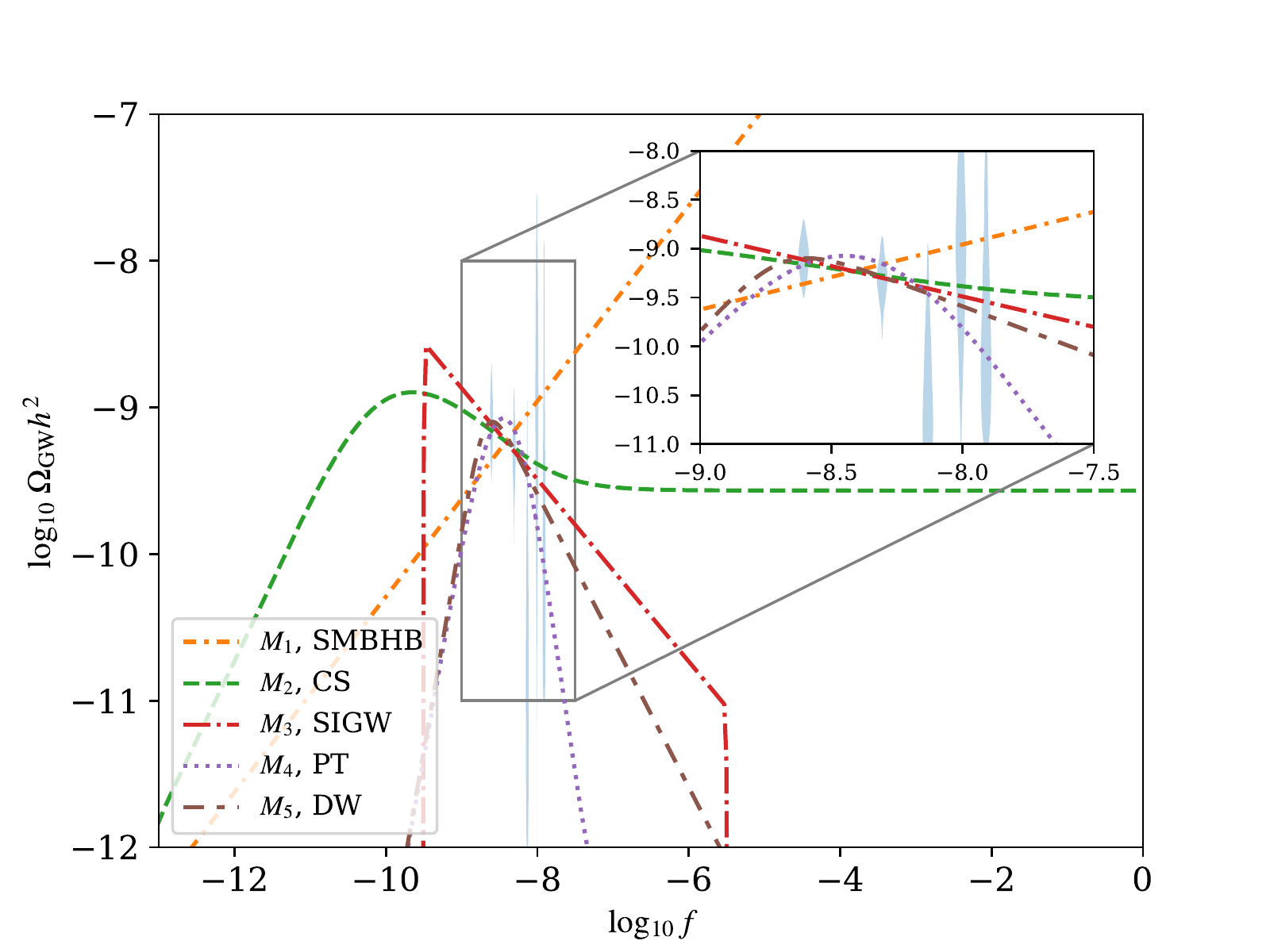}
    \includegraphics[width=0.2\textwidth]{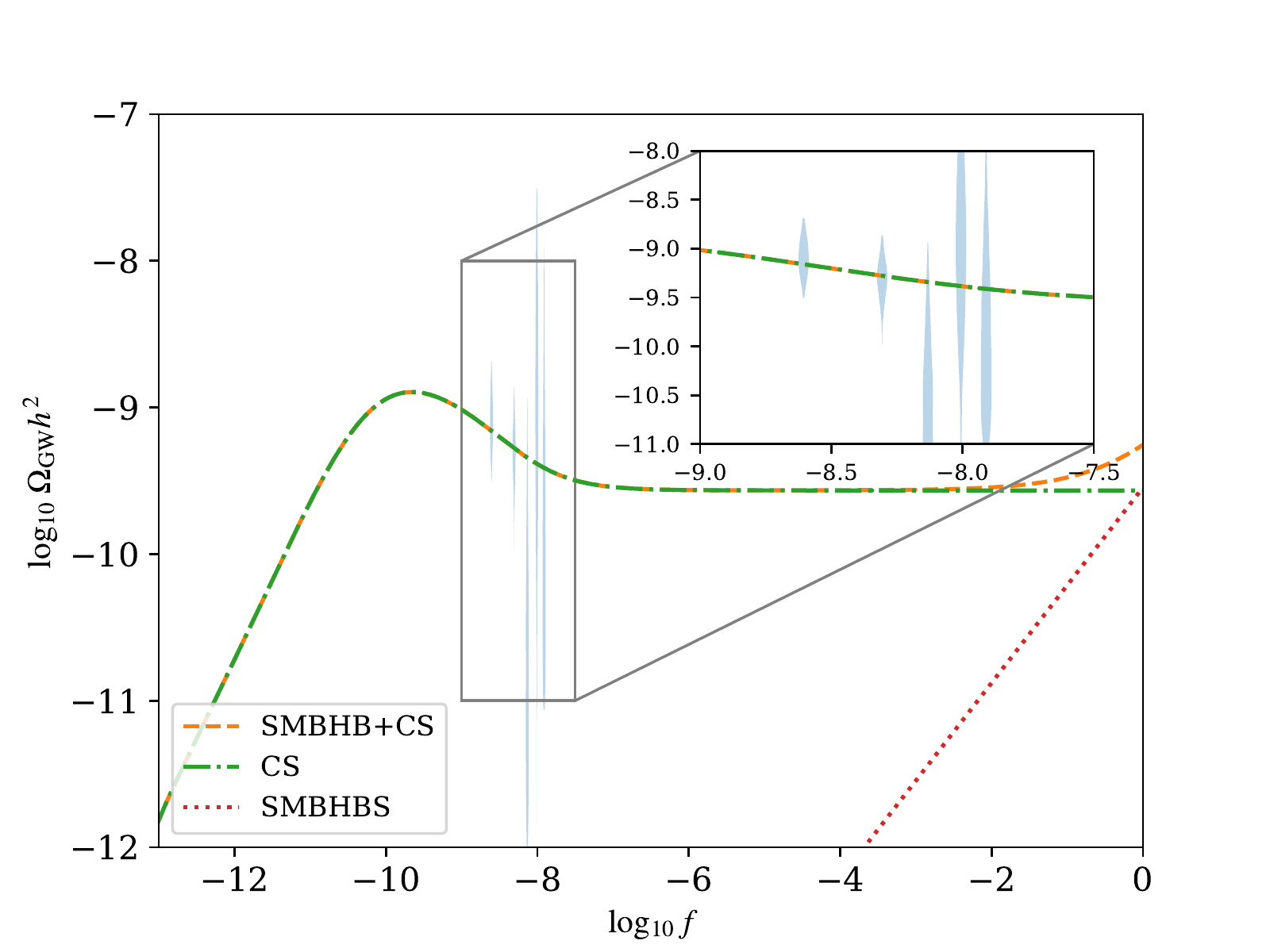}
     \includegraphics[width=0.2\textwidth]{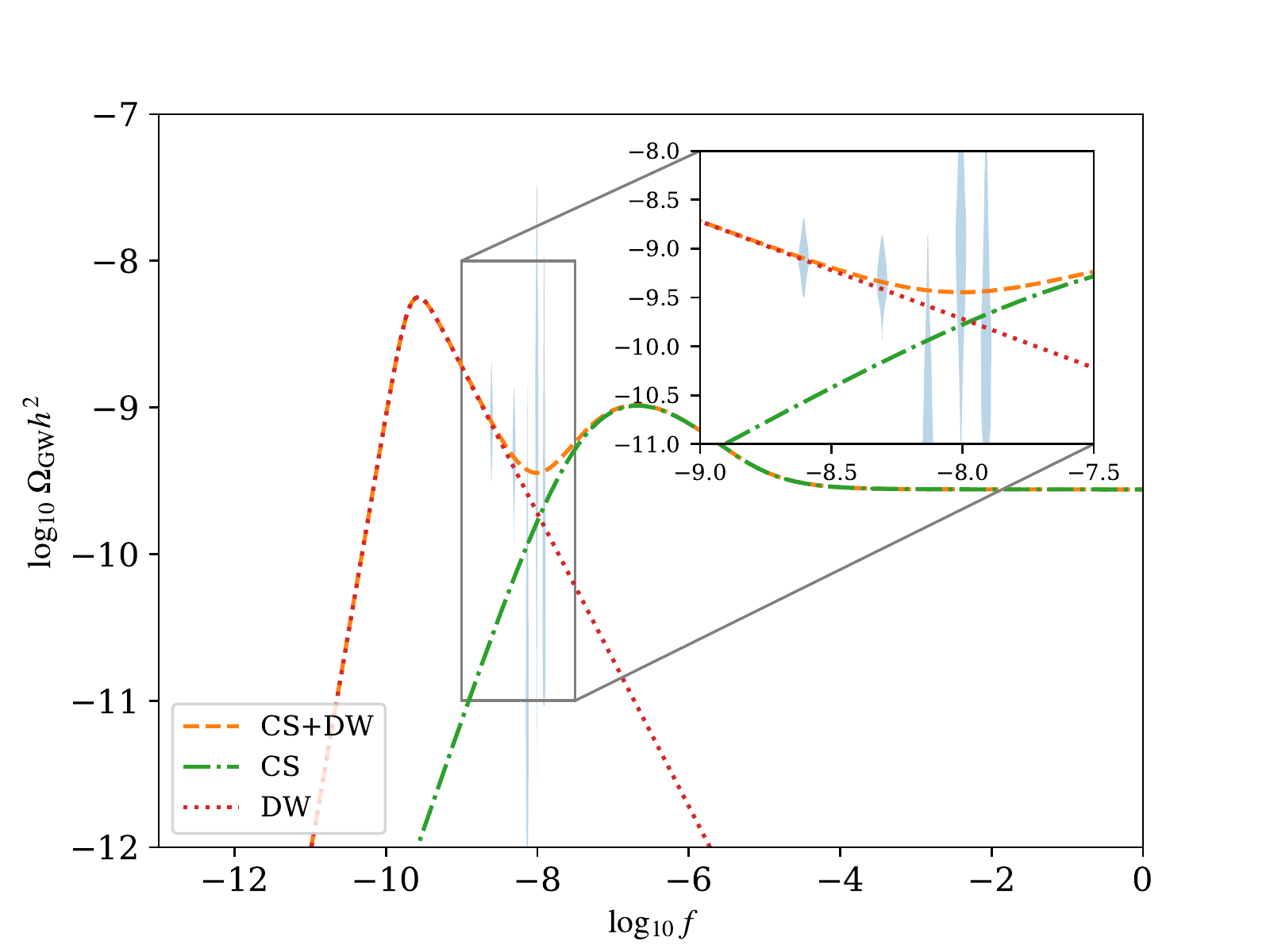}
    \includegraphics[width=0.2\textwidth]{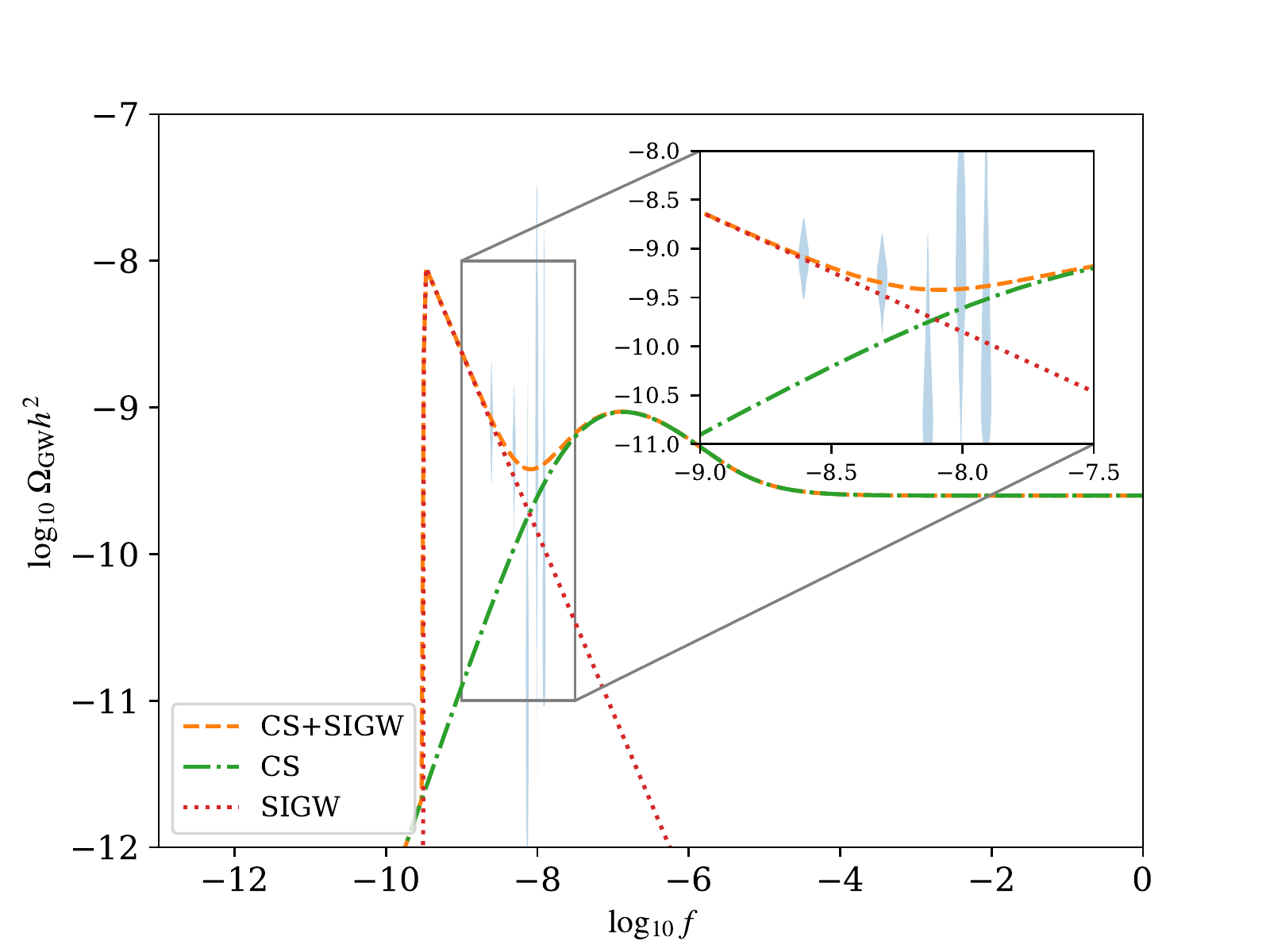}
    \caption[]{The GW energy spectrum for different models with best-fit value of parameters (which are shown in {\it supplemental materials}).
The violin plots show the first five-frequency bins of NANOGrav 12.5-yr data set. The top-left panel shows individual GW source scenario, and the other three plots
show the combined explanations with SMBHB+cosmic strings, cosmic strings+domain walls, and cosmic strings+scalar induced GWs.
    }
    \label{fig:median_fit}
\end{figure}

\noindent{\it \bfseries GW sources comparison:}
We apply Bayesian model comparison on the following models:
SMBHBs, cosmic strings, scalar induced GWs, FOPT, domain walls, SMBHBs+cosmic strings, cosmic strings + scalar induced GWs, cosmic strings + domain walls, respectively.  
 The result is given in Eq (\ref{eqB}) with the interpretation of Bayes factors are given in Table~\ref{tab:Bayes_factor_intp}~\footnote{We note that the GW signals from all these SGWB sources that can fit the NANOGrav 12.5-yr results are in tension with the previous bound of
PPTA~\cite{Shannon:2015ect} for some parameter spaces, which may be reduced when the improvement of Bayesian priors for the intrinsic pulsar red noise are adopted~\cite{Arzoumanian:2020vkk}.}. 
We find that, the current data shows a positive evidence in favor of the cosmic strings explanation against SMBHBs, scalar induced GWs,  FOPT, and domain walls, and a weak evidence in favor of cosmic strings against cosmic strings+SMBHBs, cosmic strings+scalar induced GWs, and cosmic strings+domain walls. The data shows a positive evidence of cosmic strings+SMBHBs, cosmic strings+scalar induced GWs, and cosmic string+domain walls against SMBHBs, see {\it supplemental materials} for details.
In comparison with other explanations, there is a least possibility to explain the common-spectrum process as the GWs from SMBHBs. 
The GWs spectra of different sources confront with the low frequency five bins data of NANOGrav 12.5-yr result is shown in Fig.~\ref{fig:median_fit}.  It shows that the GWs from the cosmic strings fit the data much better due to the fall-off behavior of the GW energy spectrum connecting the matter dominate region (low frequency) and radiation dominate region (high frequency).

\begin{equation}\label{eqB}
   B_{ij}= 
\begin{pmatrix}
    1 & 0.09 &0.37&0.28&0.83& 0.16 & 0.12 & 0.17\\
    10.8 & 1 & 3.96&3.01 &  8.93&1.75&1.32&1.84\\
    2.73 & 0.25 & 1 & 0.76 & 2.26&0.44&0.33&0.47\\
    3.6&  0.33&1.32 &1 & 2.97 & 0.58& 0.44 & 0.61\\
    1.21 & 0.11 & 0.44 & 0.34& 1& 0.2 & 0.15 & 0.21\\
    6.18& 0.57 & 2.26 & 1.72 & 5.11 & 1& 0.76 &1.05\\
    8.17 & 0.76 & 2.99 & 2.27 &6.75 & 1.32& 1& 1.39\\
    5.86 & 0.54 & 2.15 &1.63& 4.85 & 0.95 & 0.72 &1
\end{pmatrix}
\end{equation}

\begin{table}[htpb]
    \centering
    \caption[]{
        Bayes factors can be interpreted as follows:
        given candidate models $M_{i}$ and $M_{j}$, a Bayes factor of 20 corresponds to a belief of 95\% in the statement ``$M_{i}$ is true'', this corresponds to strong evidence in favor of $M_{i}$~\cite{10.2307/2291091}.
    }
    \label{tab:Bayes_factor_intp}
    \begin{tabular}{lr}
        \hline
        $B_{ij}$ &  Evidence in favor of $M_{i}$ against $M_{j}$ \\
        \hline
        $1-3$ & Weak \\
        $3-20$ & Positive\\
        $20-150$ & Strong\\
        $\ge 150$ & Very strong \\
        \hline
    \end{tabular}
\end{table}

\noindent{\it \bfseries Conclusion and discussion:} In this letter, we evaluate the possibility of the SGWB explanations for the stochastic common-spectrum process observed by the NANOGrav 12.5-yr data set, and perform a model comparison based on Bayesian analysis using the five low-frequency bin data. The models include GWs from SMBHBs, cosmic strings, scalar induced GWs, phase transitionan and domain walls. We also consider the situation that the SGWB is superposed by two individual sources, including cosmic strings+SMBHBs, cosmic string+scalar induced GWs, and cosmic strings+domain walls.  We evaluate the possibility of these SGWB explanations for the stochastic common-spectrum process observed by the NANOGrav 12.5-yr data set, the analysis shows that:  1) with a positive evidence, the cosmic string model is the most favored one by the current data against SMBHBs and/or other SGWB sources, which hints that the symmetry breaking scale of a $U(1)$ symmetry of new physics close to the symmetry breaking scale of GUT~\cite{Buchmuller:2019gfy,Dror:2019syi,King:2020hyd}; 2) a lot parameter spaces for the explanation of the dark sector FOPT is invalidated by the BBN bounds;  3) scalar-induced GWs hints the mass of PBHs around solar mass and its abundance as dark matter. 
Other combinations of GWs sources are less supported by data.

Refs.~\cite{Ellis:2020ena,Blasi:2020mfx} study cosmic strings as the possible SGWB sources of the NANOGrav data, both the two studies are performed based on a power-law spectrum. Ref.~\cite{Ellis:2020ena} assumes a single $\alpha_{CS}=0.1$, and finds $G\mu\sim [4,10]\times 10^{-11}$ at 1 $\sigma$. 
While Ref.~\cite{Blasi:2020mfx}
assumes $\alpha_{CS}$ and $G\mu$ to be free parameters  and obtains $G\mu\sim [6,17]\times 10^{-11}$ and $\alpha_{CS}\sim [10^{-2},10^{-1}]$ at 1 $\sigma$. Our study is based on the Bayesian analysis of NANOGrav data disregarding the power-law spectrum. We show that the NANOGrav 12.5-yr data favors a wider range of $G\mu$ and a smaller $\alpha_{CS}$ at 1 $\sigma$, which corresponds to a spontaneous breaking scale of a local $U(1)$ symmetry closer to the Grant Unification scale in comparison with the above two papers, whose GW energy spectrum may be checked in high frequency regions by LIGO, LISA, Tianqin, and Taiji\footnote{At last, we comment that
the fragmentation of the inflaton could product GWs during preheating process~\cite{Khlebnikov:1997di,Dufaux:2007pt}. At the end of inflation, the inflaton starts to oscillate around the minimum of the effective potential, and the perturbations of the inflaton are exponentially amplified due to parametric resonance. The peak value of $\Omega_{\mathrm{GW}}$ comes within $10^{-9}$ to $10^{-11}$. However, since the peak frequency of the SGWB is proportional to the inflationary energy scale~\cite{Easther:2006vd}, to generate a SGWB with peak frequency $10^{-9}$ Hz, the inflationary energy scale is below $100$ MeV, which is too low to reheat the universe. Ref.~\cite{Vagnozzi:2020gtf} shows that the inflationary interpretation of the NANOGrav result is impossible as we comment here. }.

\noindent{\it \bfseries Acknowledgements:}
We are grateful to Xiao Xue, Qiang Yuan, and Xingjiang Zhu for helpful discussions.
RGCai was supported by the National Natural Science Foundation of China Grants No.11947302, No.11991052, No.11690022, No.11821505 and No.11851302. Ligong Bian was supported by the National Natural Science Foundation of China Grants No.12075041, No.11605016, No.11947406, No.12047564, the Fundamental Research Funds for the Central Universities under Grant No. 2020CDJQY-Z003, and Chongqing Natural Science Foundation under Grant No.cstc2020jcyj-msxmX0814.

\newpage

\appendix

\section{GWs formula for Scalar-induced GWs}

The energy spectrum $\Omega_{\mathrm{GW}}$ observed at $t_{0}$ can be expressed in terms of the power spectrum of scalar perturbations $P_\mathcal{R}(k)$ as~\cite{Kohri:2018awv}
\begin{equation}
	\begin{split}
		\Omega_{\mathrm{GW}}^{si}(f)h^{2}
		&= \dfrac{1}{12}\Omega_{\mathrm{rad}}h^{2}\left(\dfrac{g_{0}}{g_{*}}\right)^{\frac{1}{3}}\times\\
		&\int_{0}^{\infty} dv \int_{|1-v|}^{1+v} du \left(\frac{4 v^{2}-\left(1+v^{2}-u^{2}\right)^{2}}{4 u v}\right)^{2}\times\\
		&
		P_{\mathcal{R}} (2\pi f u) P_{\mathcal{R}} (2\pi f v) I^{2}(u,v)\,,
	\end{split}
	\label{eq:final}
\end{equation}
where $g_{0}$ and $g_{*}$ are the effective relativistic degrees of freedom at $t_{0}$ and at the time when the $k$-mode crosses the Hubble horizon, $\Omega_{\mathrm{rad}}h^{2}=4.2\times 10^{-5}$ is the density fraction of radiation at $t_{0}$, and
\begin{equation}
	\begin{split}
		I^{2}(u,v)
		&= \dfrac{1}{2}\left(\frac{3}{4 u^{3} v^{3} x}\right)^{2}\left(u^{2}+v^{2}-3\right)^{2} \\
		&\times\Bigg\{ \left[-4 u v+\left(u^{2}+v^{2}-3\right) \ln \left|\frac{3-(u+v)^{2}}{3-(u-v)^{2}}\right|\right]^{2} \\
		&  +\left[\pi\left(u^{2}+v^{2}-3\right) \Theta(u+v-\sqrt{3})\right]^{2} \Bigg\}\,.
	\end{split}
	\label{eq:I3}
\end{equation}

Consider the case $P_{\mathcal{R}}(k)$ has a power-law form around $k_{*}=\frac{2\pi}{1\mathrm{yr}}=20.6\mathrm{pc}^{-1}$,
\begin{equation}
	\label{eq:PR}
	P_{\mathcal{R}}(k)=P_{\mathcal{R}0}\left(\dfrac{k}{k_{*}}\right)^{m}\Theta(k-k_{min})\Theta(k_{max}-k)\,,
\end{equation}
where $\Theta(x)$ is the Heaviside function and $k_{*}=\frac{2\pi}{1\mathrm{yr}}=20.6\mathrm{pc}^{-1}$.
To prevent overproduction of PBHs, we set cutoffs at $k_{min}=0.03k_{*}$ and $k_{max}=100k_{*}$ so that $P_{\mathcal{R}}(k)$ obtains an upper bound.

\section{Bayesian model fitting and model comparison}

$D$, $M$ and $\theta$ denote data, model and parameters of model, respectively.

Bayesian model fitting: finding the posterior of parameters when data and model are known.
The Bayes' theorem gives
\begin{equation}
    P(\theta|D,M) = \frac{P(D|\theta,M) P(\theta|M)}{P(D|M)},
\end{equation}
where $P(\theta|D,M)$, $P(D|\theta,M)$,  $P(\theta|M)$ and $P(D|M)$ are called posterior, likelihood, prior and model evidence, respectively. 

Bayesian model comparison: finding the model evidence when data are known.
The Bayes' theorem gives
\begin{equation}
    P(M|D)= P(D|M) \frac{P(M)}{P(D)}.
\end{equation}
Using the definition of conditional probability, $P(D|M)$ can be expressed as an integral over the parameter space of the likelihood,
\begin{equation}
    P(D|M)=\int P(D|\theta,M) P(\theta|M) d\theta.
\end{equation}
Then the odds ratio between two alternative models is
\begin{equation}
    O_{ij} \equiv \frac{P(M_{i}|D)}{P(M_{j}|D)} = \frac{P(D|M_{i})}{P(D|M_{j})} \frac{P(M_{i})}{P(M_{j})} = B_{ij} \frac{P(M_{i})}{P(M_{j})} 
\end{equation}
where $B_{ij} \equiv P(D|M_{i})/P(D|M_{j})$ is the Bayes factor for $M_{i}$ versus $M_{j}$, and $P(M_{i})/P(M_{j})$ is the ratio of prior odds which is usually assumed to be near unity.

\section{Results of Bayesian model fitting on NANOGrav 12.5-yr data}

\begin{figure}[!htp]
    \centering
    \includegraphics[width=0.4\textwidth]{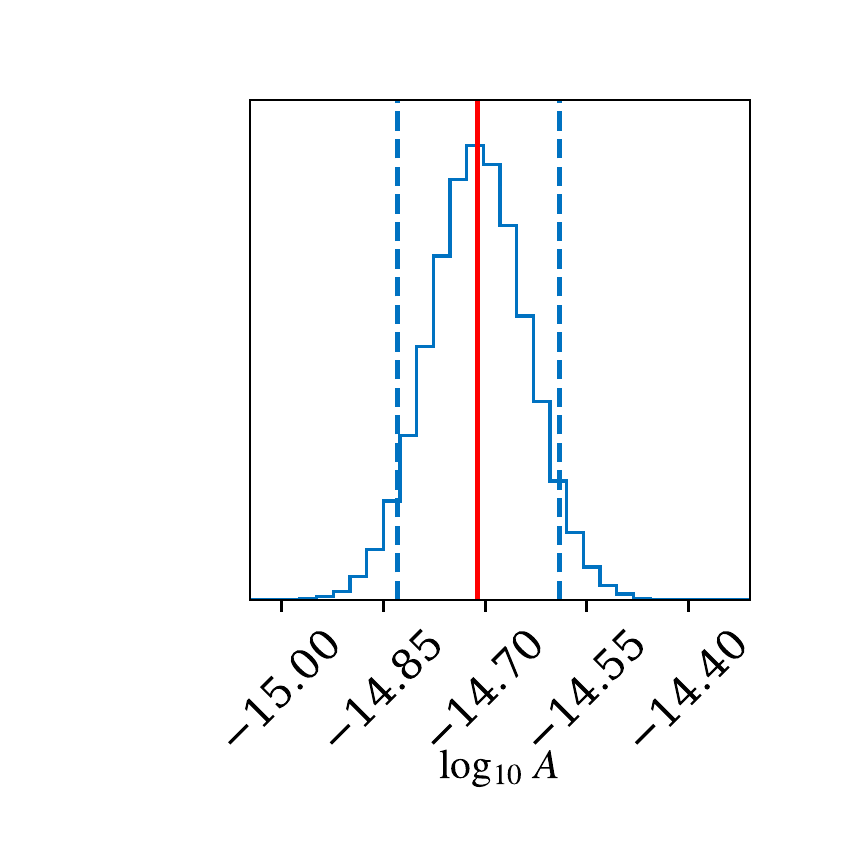}
    \caption[]{{$\bullet$ \bf Super massive black hole binaries (SMBHBs).} Triangle plots of posteriors given by fitting the first five frequency bins of NANOGrav 12.5-yr data set.
        The red solid lines indicate the best-fit parameter values, and the blue dashed lines indicate the 5\% and 95\% quantiles for each parameter.
               $\log_{10} A$ has best-fit $-14.71$ and $5\% \sim 95\%$ quantiles of $-14.83\sim-14.59$. 
    }
    \label{fig:corner_M1}
\end{figure}

\begin{figure}[!htp]
    \centering
    \includegraphics[width=0.4\textwidth]{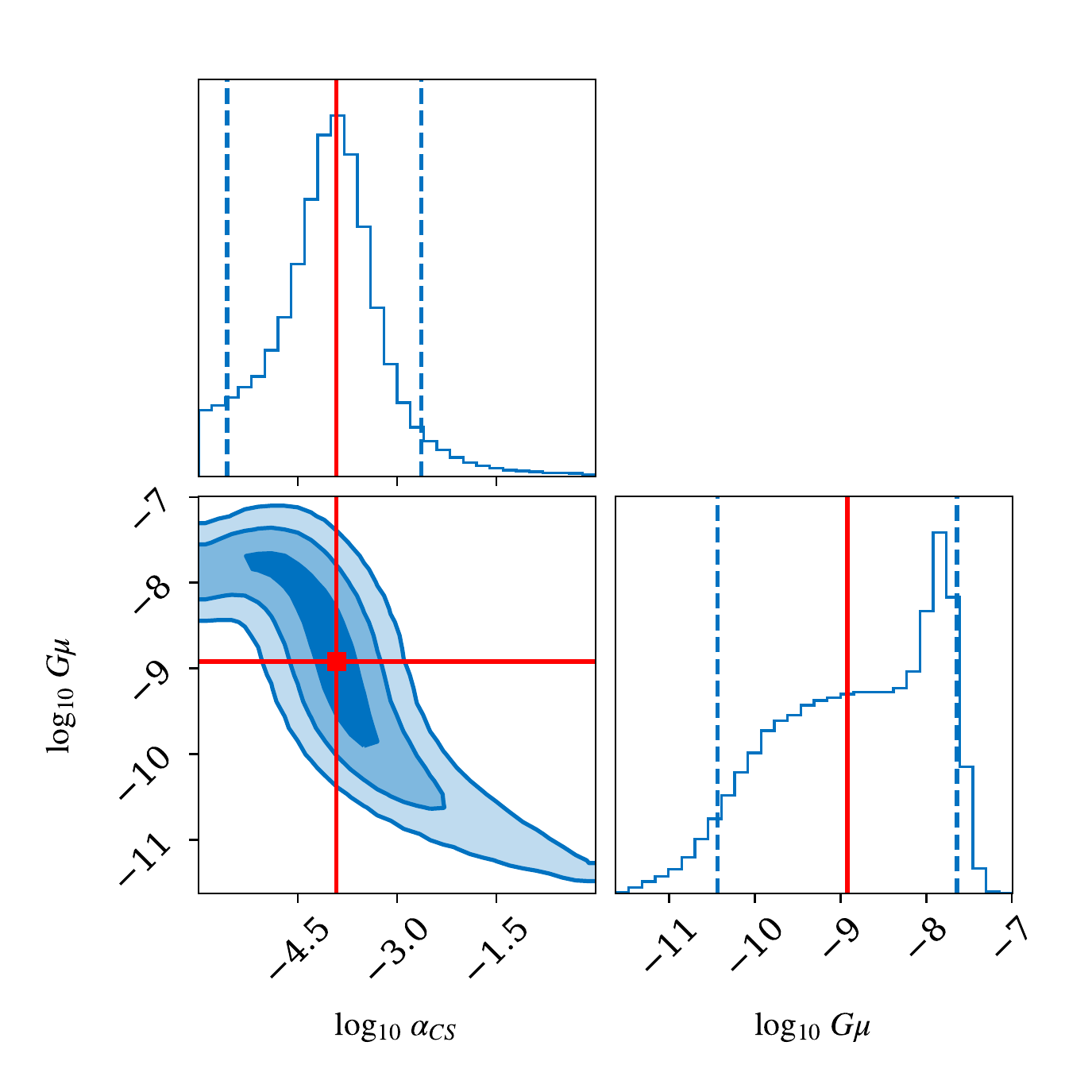}
    \caption[]{$\bullet${\bf  Cosmic strings. }Triangle plots of posteriors given by fitting the first five frequency bins of NANOGrav 12.5-yr data set.
        Three different contour levels in the plots are 0.39, 0.86 and 0.99, which corresponds to 1-, 2- and 3-$\sigma$ level in 2D distributions.
        The red solid lines indicate the best-fit parameter values, and the blue dashed lines indicate the 5\% and 95\% quantiles for each parameter.
        $\log_{10} \alpha_{CS}$ has best-fit
         $-3.92$ and $5\% \sim 95\%$ quantiles of $-5.57\sim -2.63$, $\log_{10} G\mu$ has best-fit $-8.92$ and $5\% \sim 95\%$ quantiles of $-10.44 \sim -7.64$.
    }
    \label{fig:corner_M2}
\end{figure}

\begin{figure}[htpb]
    \centering
    \includegraphics[width=0.4\textwidth]{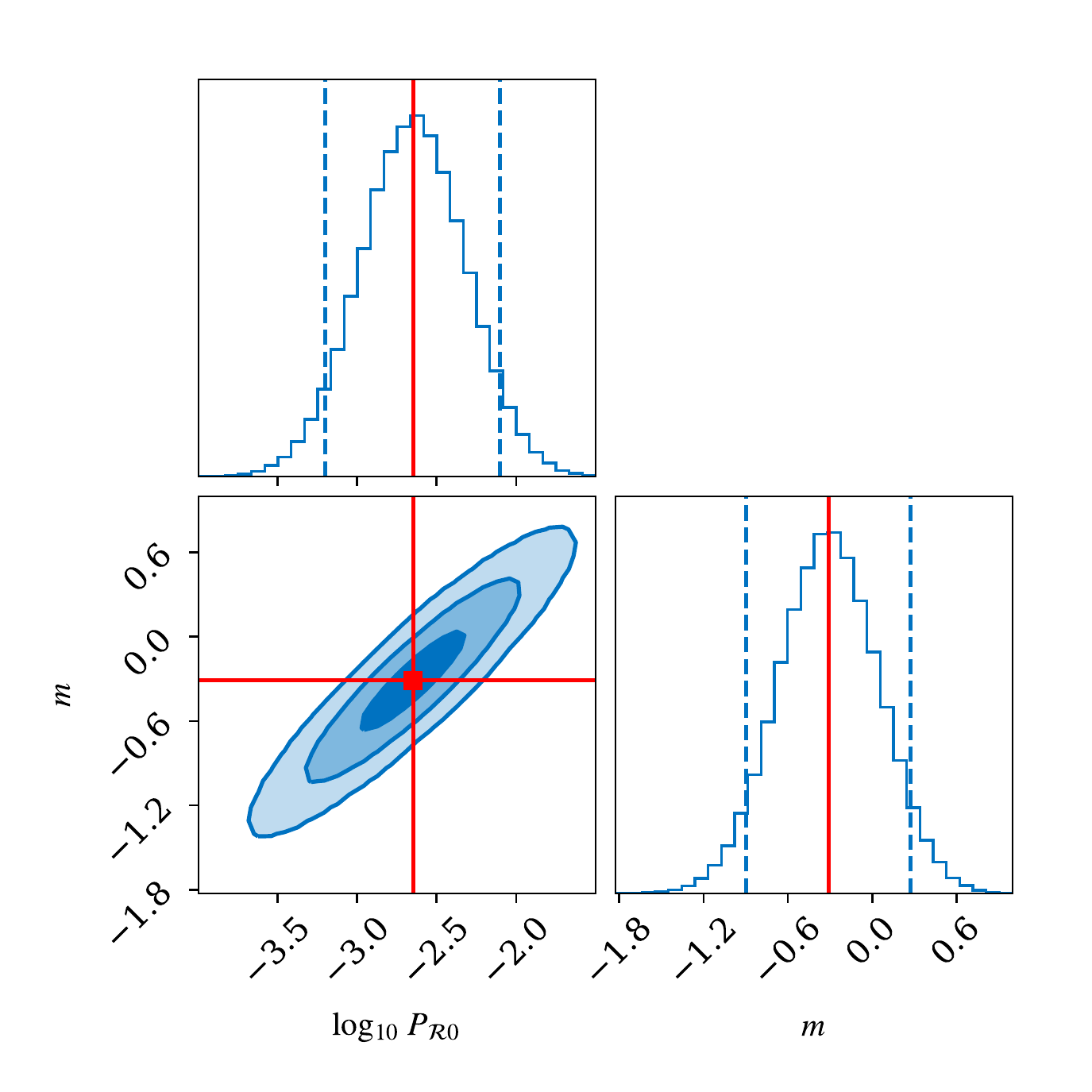}
       \caption[]{ 
$\bullet${\bf  Scalar induced GW.}
     Triangle plots of posteriors given by fitting the first five frequency bins of NANOGrav 12.5-yr data set.
        Three different contour levels in the plots are 0.39, 0.86 and 0.99, which corresponds to 1-, 2- and 3-$\sigma$ level in 2D distributions.
        The red solid lines indicate the best-fit parameter values, and the blue dashed lines indicate the 5\% and 95\% quantiles for each parameter. $\log_{10} P_{\mathcal{R}0}$ has best-fit $-2.65$ and $5\% \sim 95\%$ quantiles of $-3.20 \sim -2.10$,
        $m$ has best-fit $-0.31$ and $5\% \sim 95\%$ quantiles of $-0.89 \sim 0.27$.
    }
    \label{fig:corner_M3}
\end{figure}

\begin{figure}[htpb]
    \centering
    \includegraphics[width=0.4\textwidth]{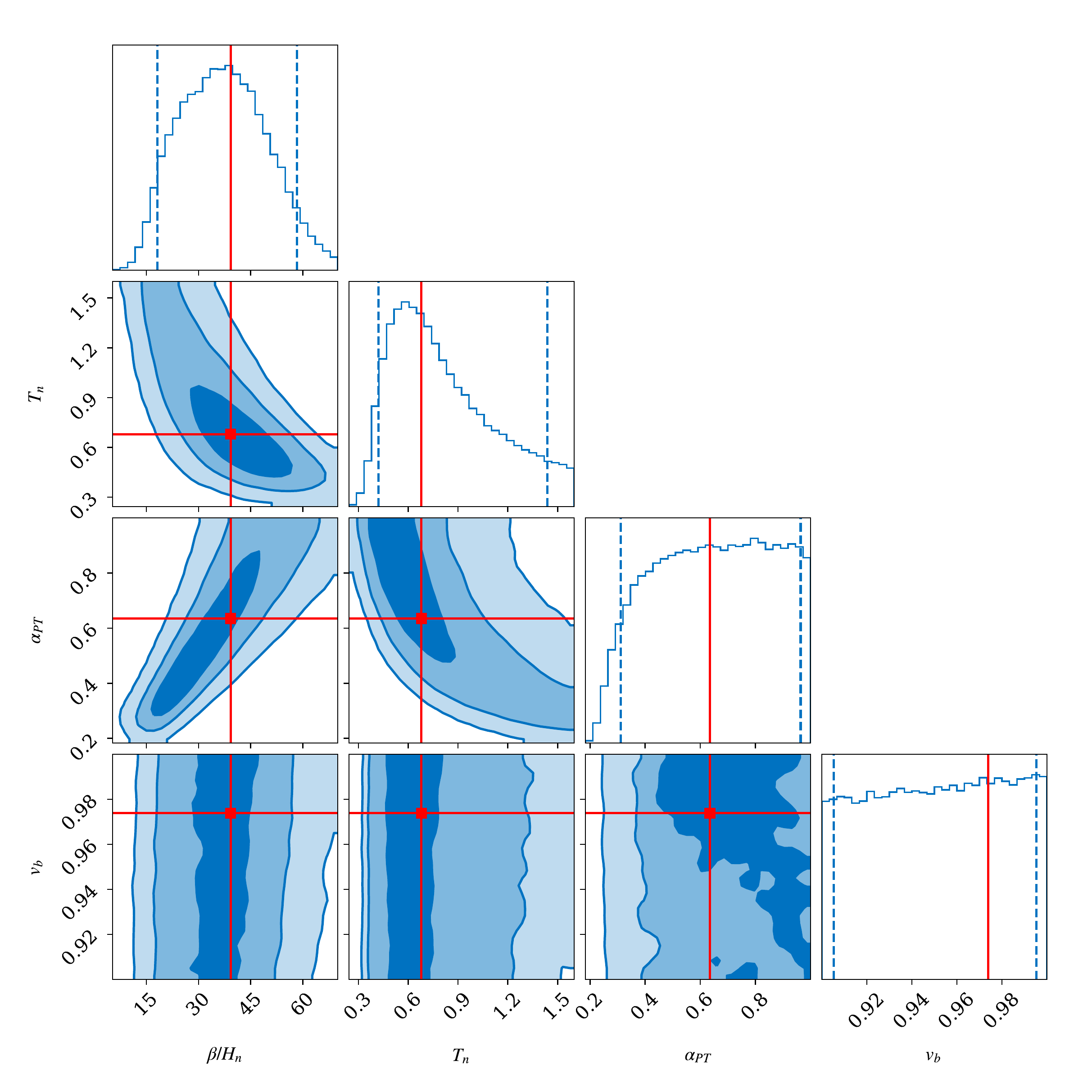}
    \caption[]{$\bullet${ \bf FOPT.} Triangle plots of posteriors given by fitting the first five frequency bins of NANOGrav 12.5-yr data set.
        Three different contour levels in the plots are 0.39, 0.86 and 0.99, which corresponds to 1-, 2- and 3-$\sigma$ level in 2D distributions.
        The red solid lines indicate the best-fit parameter values, and the blue dashed lines indicate the 5\% and 95\% quantiles for each parameter.
       $T_n$ has best-fit
         $0.68$ and $5\% \sim 95\%$ quantiles of $0.42\sim 1.43$, $\alpha_{PT}$ has best-fit
         $0.64$ and $5\% \sim 95\%$ quantiles of $0.31\sim0.96$, $\beta/H_{n}$ has best-fit
         $39.22$ and $5\% \sim 95\%$ quantiles of $18.17\sim58.27$, $v_b$ has best-fit
         $0.97$ and $5\% \sim 95\%$ quantiles of $0.91\sim0.99$.
           }
    \label{fig:corner_M4}
\end{figure}

\begin{figure}[htpb]
    \centering
    \includegraphics[width=0.4\textwidth]{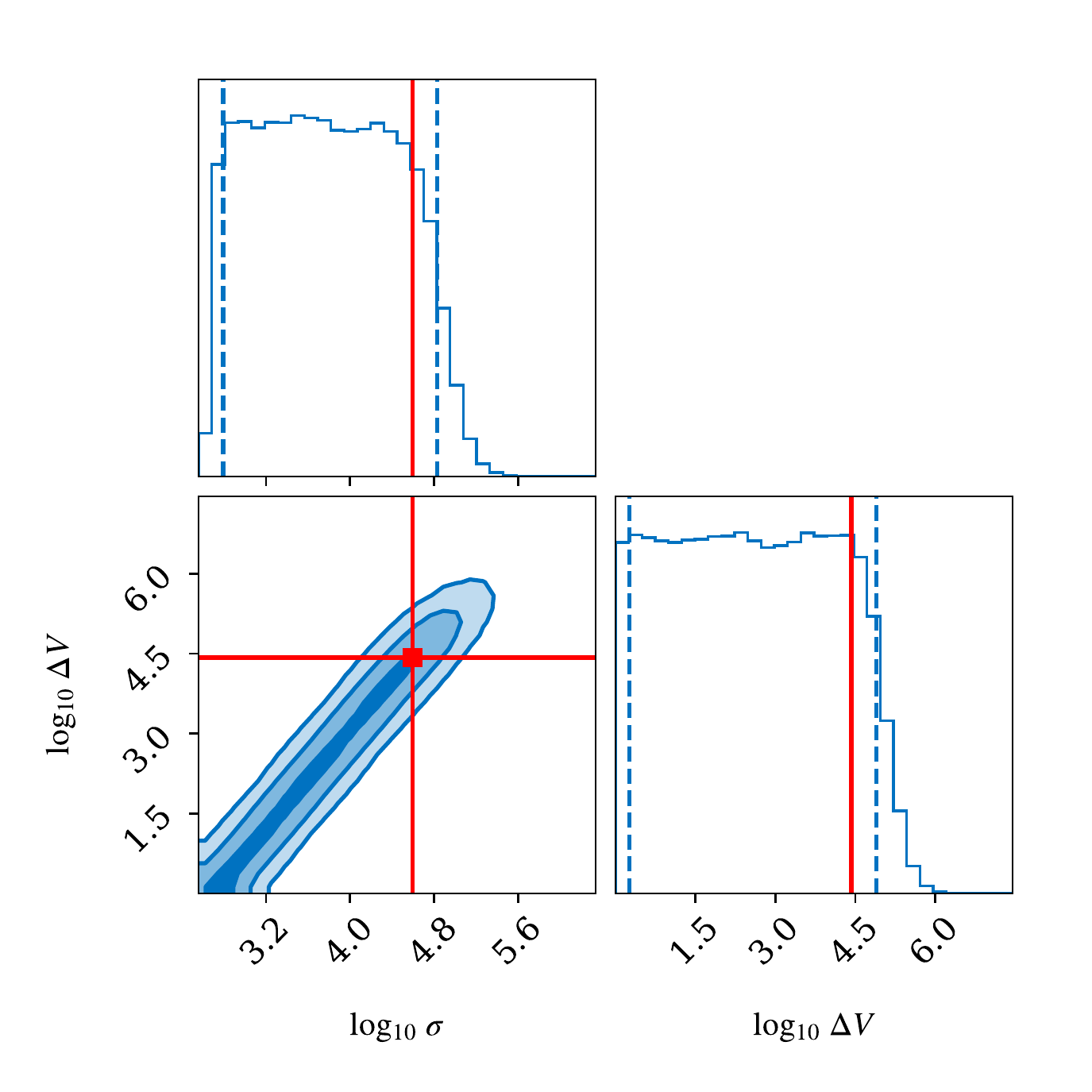}
    \caption[]{$\bullet$ { \bf Domain walls.}
Triangle plots of posteriors given by fitting the first five frequency bins of NANOGrav 12.5-yr data set.
        Three different contour levels in the plots are 0.39, 0.86 and 0.99, which corresponds to 1-, 2- and 3-$\sigma$ level in 2D distributions.
        The red solid lines indicate the best-fit parameter values, and the blue dashed lines indicate the 5\% and 95\% quantiles for each parameter.
        $\log_{10} (\sigma/\mathrm{TeV}^{3})$ has best-fit $4.59$ and $5\% \sim 95\%$ quantiles of $2.79\sim 4.83$,
        $\log_{10} (\Delta V/\mathrm{MeV}^{4})$ has best-fit $4.43$ and $5\% \sim 95\%$ quantiles of $0.26 \sim 4.80$.   
        }
    \label{fig:corner_M5}
\end{figure}

\begin{figure}[htpb]
    \centering
    \includegraphics[width=0.4\textwidth]{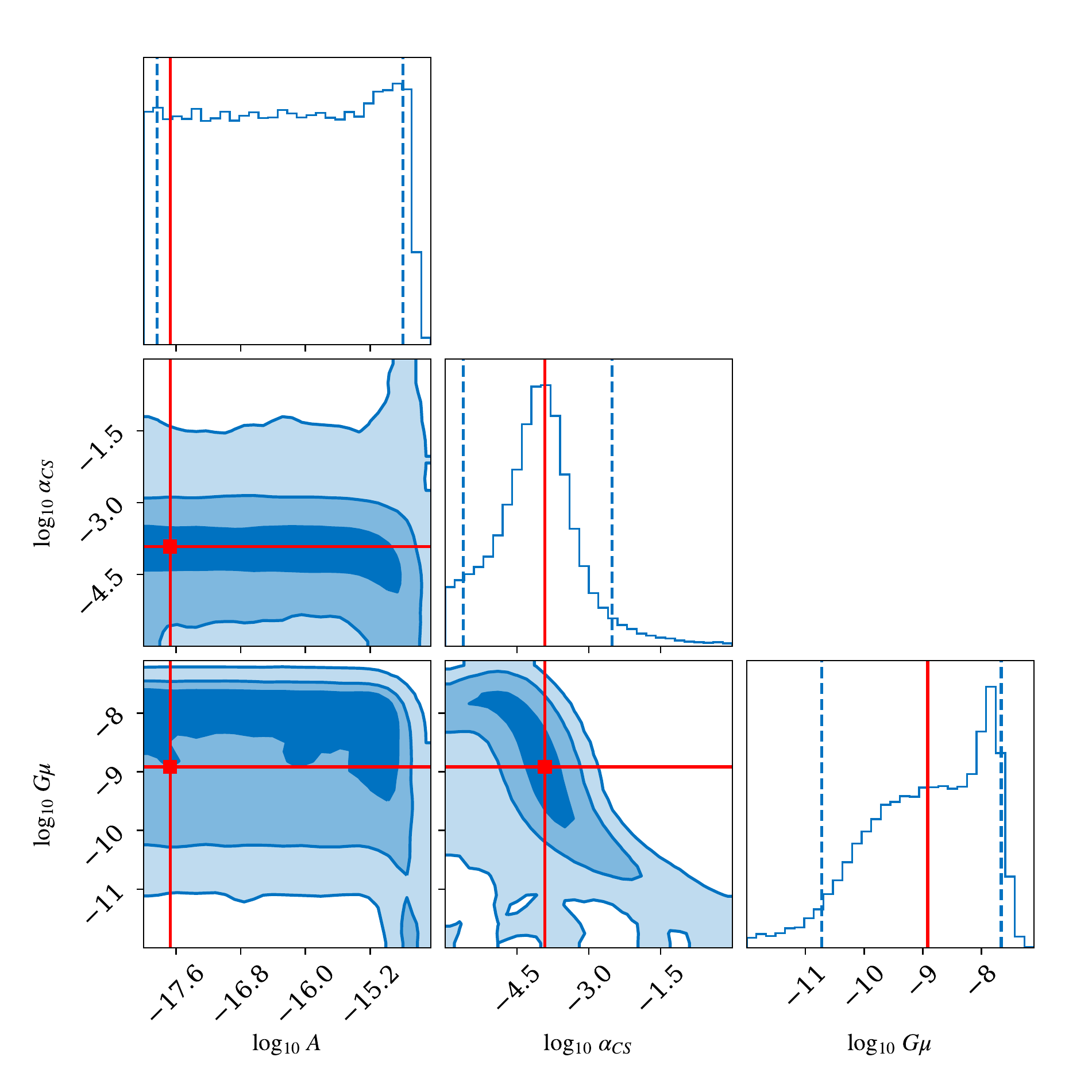}
    \caption[]{$\bullet$ {\bf SMBHBs+ Cosmic strings.} Triangle plots of posteriors given by fitting the first five frequency bins of NANOGrav 12.5-yr data set.
        Three different contour levels in the plots are 0.39, 0.86 and 0.99, which corresponds to 1-, 2- and 3-$\sigma$ level in 2D distributions.
        The red solid lines indicate the best-fit parameter values, and the blue dashed lines indicate the 5\% and 95\% quantiles for each parameter.
        $\log_{10} \alpha_{CS}$ has best-fit
         $-3.92$ and $5\% \sim 95\%$ quantiles of $-5.62\sim -2.52$, $\log_{10} G\mu$ has best-fit $-8.91$ and $5\% \sim 95\%$ quantiles of $-10.72 \sim -7.66$, $\log_{10} A$ has best-fit $-17.67$ and $5\% \sim 95\%$ quantiles of $-17.83\sim-14.79$.
    }
    \label{fig:corner_M6}
\end{figure}

\begin{figure}[htpb]
    \centering
    \includegraphics[width=0.4\textwidth]{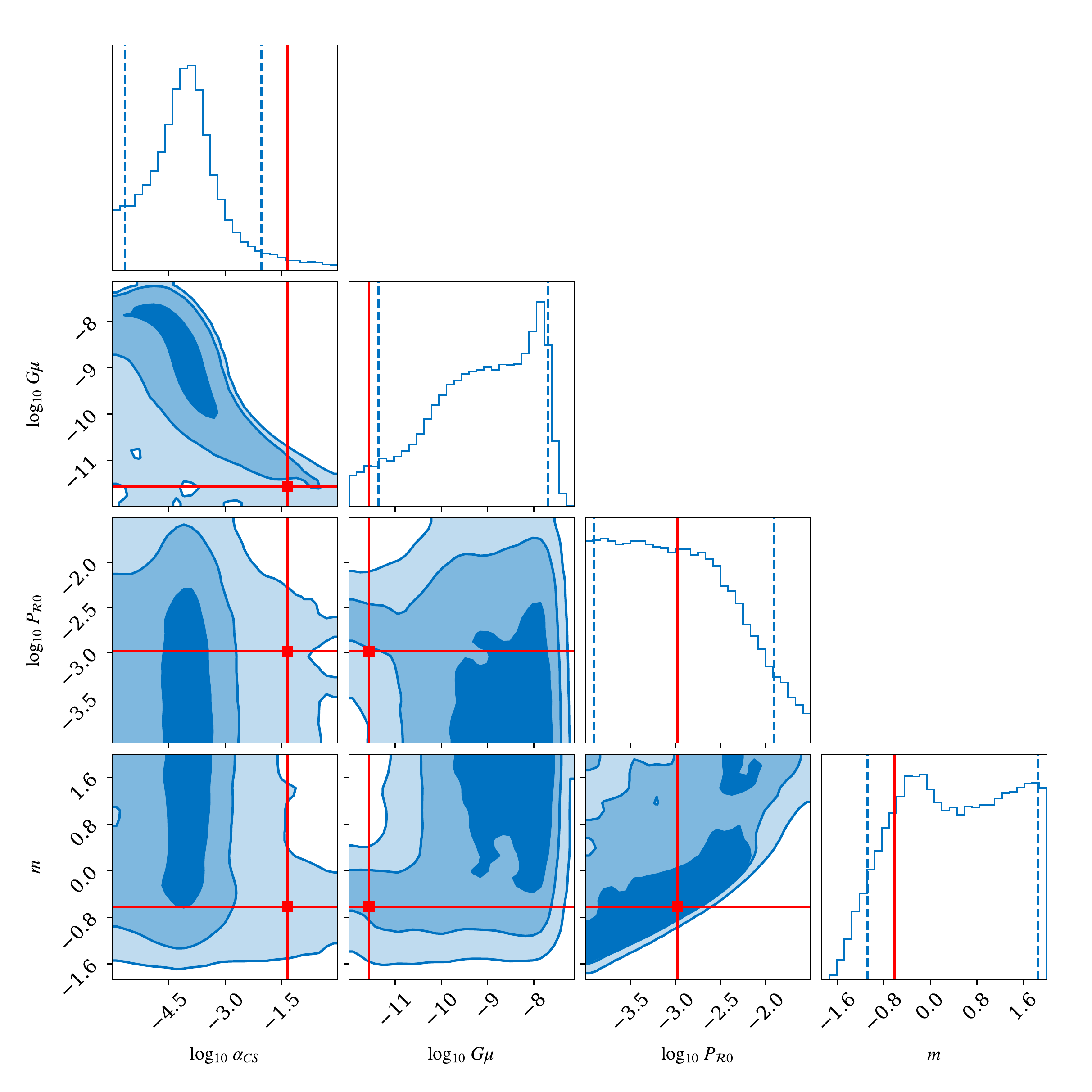}
    \caption[]{$\bullet${ \bf Cosmic strings + Scalar induced GWs.} Triangle plots of posteriors given by fitting the first five frequency bins of NANOGrav 12.5-yr data set.
        Three different contour levels in the plots are 0.39, 0.86 and 0.99, which corresponds to 1-, 2- and 3-$\sigma$ level in 2D distributions.
        The red solid lines indicate the best-fit parameter values, and the blue dashed lines indicate the 5\% and 95\% quantiles for each parameter.
$\log_{10} \alpha_{CS}$ has best-fit
         $   -1.33$ and $5\% \sim 95\%$ quantiles of $-5.67\sim -2.03$, $\log_{10} G\mu$ has best-fit $-11.57
$ and $5\% \sim 95\%$ quantiles of $-11.35 \sim -7.69$,       
        $\log_{10} P_{\mathcal{R}0}$ has best-fit $-2.98$ and $5\% \sim 95\%$ quantiles of $-3.90 \sim -1.91$,
        $m$ has best-fit $-0.61$ and $5\% \sim 95\%$ quantiles of $-1.08 \sim 1.85$.
    }
    \label{fig:corner_M7}
\end{figure}

\begin{figure}[htpb]
    \centering
    \includegraphics[width=0.4\textwidth]{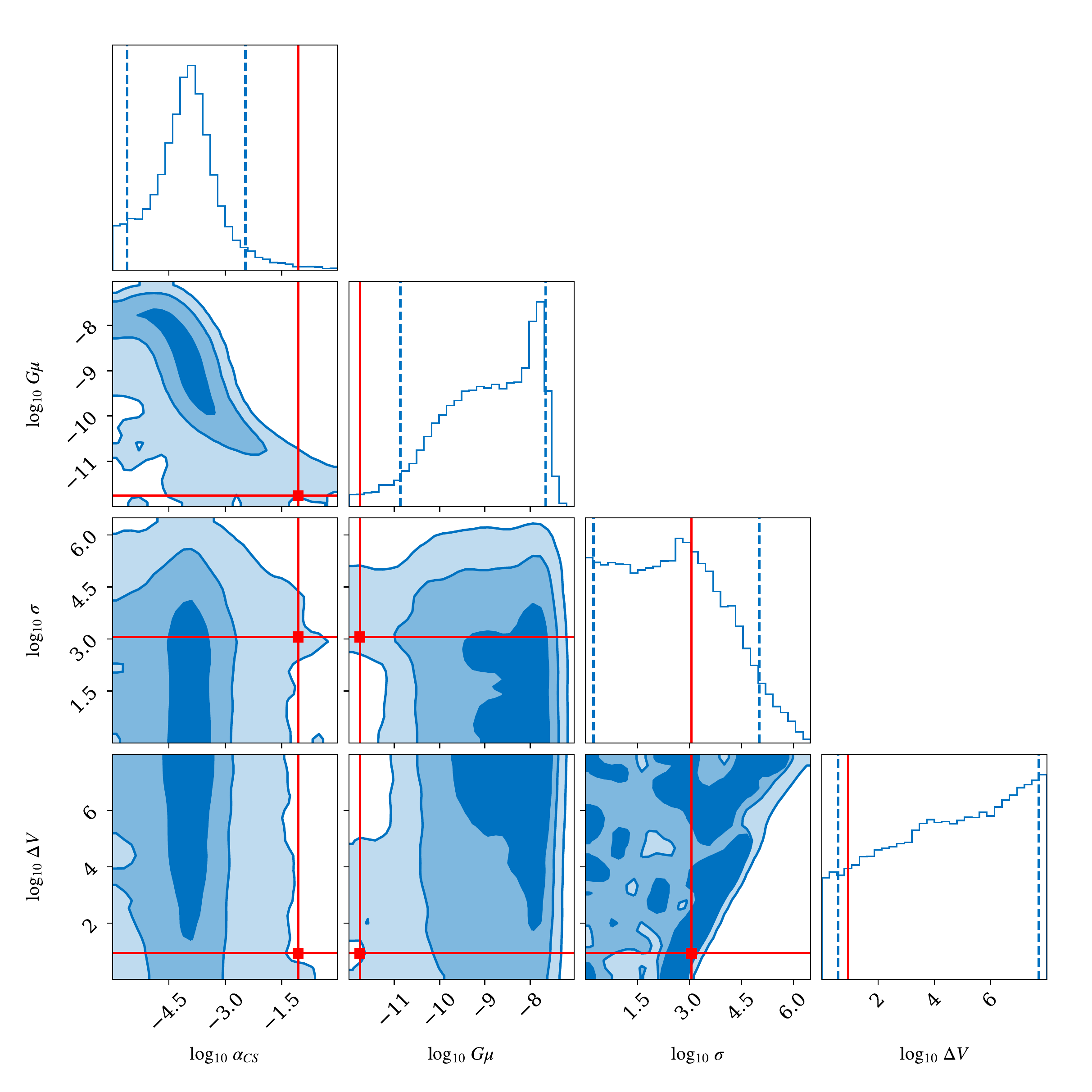}
    \caption[]{$\bullet${ \bf  Cosmic strings+ Domain walls.} Triangle plots of posteriors given by fitting the first five frequency bins of NANOGrav 12.5-yr data set.
        Three different contour levels in the plots are 0.39, 0.86 and 0.99, which corresponds to 1-, 2- and 3-$\sigma$ level in 2D distributions.
        The red solid lines indicate the best-fit parameter values, and the blue dashed lines indicate the 5\% and 95\% quantiles for each parameter.
$\log_{10} \alpha_{CS}$ has best-fit
         $   -1.06$ and $5\% \sim 95\%$ quantiles of $-5.61\sim -2.46$, $\log_{10} G\mu$ has best-fit $
-11.76$ and $5\% \sim 95\%$ quantiles of $-10.86 \sim -7.66$,    
$\log_{10} (\sigma/\mathrm{TeV}^{3})$ has best-fit $3.06$ and $5\% \sim 95\%$ quantiles of $0.23\sim 5.01$,
        $\log_{10} (\Delta V/\mathrm{MeV}^{4})$ has best-fit $0.93$ and $5\% \sim 95\%$ quantiles of $0.58 \sim 7.70$.
    }
    \label{fig:corner_M8}
\end{figure}

We present the results of Bayesian model fitting on the first five frequency bins of NANOGrav 12.5-yr data set for GW models: SMBHBs, SMBHBs+cosmic strings, cosmic strings + scalar induced GWs, cosmic strings+ domain walls, respectively. Fig.~\ref{fig:corner_M1},\ref{fig:corner_M2},\ref{fig:corner_M3},\ref{fig:corner_M4},\ref{fig:corner_M5},\ref{fig:corner_M6},\ref{fig:corner_M7},\ref{fig:corner_M8} are generated by \texttt{emcee} and \texttt{corner.py}~\cite{Foreman_Mackey_2013,corner}.

\bibliographystyle{unsrt}

\end{document}